\documentclass[twocolumn,prx,amsmath,amssymb]{revtex4}

\usepackage{graphicx}
\usepackage{float}
\usepackage{amssymb,amsthm,amsfonts,amstext}
\usepackage{url}
\usepackage{xcolor}
\usepackage{changes}
\def\be{\begin{equation}}
\def\ee{\end{equation}}
\def\bea{\begin{eqnarray}}
\def\eea{\end{eqnarray}}
\def\bma{\begin{mathletters}}
\def\ema{\end{mathletters}}

\def\q0{\underline{0}}

\def\H{{\cal H}}

\def\C{{\mathbb C}}
\def\id{{\mathbb I}}

\def\H{{\cal H}}

\def\R{\mathbb{R}}

\def\tr{\mbox{tr}}
\def\one{\leavevmode\hbox{\small1\normalsize\kern-.33em1}}

\def\bra#1{\langle#1|} \def\ket#1{|#1\rangle}

\def\proj#1{\ket{#1}\!\bra{#1}}

\newtheorem{theo}{Theorem}

\newtheorem{defin}[theo]{Definition}

\newtheorem{prop}[theo]{Proposition}

\def\id{{\mathbb I}}

\def\eref#1{Eq.~(\ref{#1})}
\def\fref#1{Fig.~\ref{#1}}

\begin{document}

\title{Connector tensor networks:\\a renormalization-type approach to quantum certification}
\author{Miguel Navascu\'es$^1$, Sukhbinder Singh$^2$ and Antonio Ac\'in$^{3,4}$}

\affiliation{$^1$Institute for Quantum Optics and Quantum Information (IQOQI) Vienna\\ Austrian Academy of Sciences\\
$^2$Max-Planck Institute for Gravitational Physics (Albert Einstein Institute), Potsdam, Germany\\
$^3$ICFO-Institut de Ci\'encies Fot\'oniques, The Barcelona Institute of Science and Technology, 08860 Castelldefels (Barcelona), Spain\\
$^4$ICREA-Instituci\`o Catalana de Recerca i Estudis Avançats, Lluis Companys 23, 08010 Barcelona, Spain}

\begin{abstract}
As quantum technologies develop, we acquire control of an ever-growing number of quantum systems. Unfortunately, current tools to detect relevant quantum properties of quantum states, such as entanglement and Bell nonlocality, suffer from severe scalability issues and can only be computed for systems of a very modest size, of around $6$ sites. In order to address large many-body systems, we propose a renormalisation-type approach based on a class of local linear transformations, called \textit{connectors}, which can be used to \textit{coarse-grain} the system in a way that preserves the property under investigation. Repeated coarse-graining produces a system of manageable size, whose properties can then be explored by means of usual techniques for small systems. In case of a successful detection of the desired property, the method outputs a linear witness which admits an exact \textit{tensor network} representation, composed of connectors. We demonstrate the power of our method by certifying using a normal desktop computer entanglement, Bell nonlocality and supra-quantum Bell nonlocality in systems with hundreds of sites. 

\end{abstract}

\maketitle
\tableofcontents
\section{Introduction}

A central goal in quantum information theory is to detect interesting global properties of few or many-body systems. For example, traditionally one may be interested in detecting whether a given quantum many-body state is entangled \cite{reviewEnt}, or whether a given conditional probability distribution contains non-classical correlations, in the sense of violating a Bell inequality~\cite{reviewNL}. More recent ventures include detecting quantum causality properties \cite{quantNetwork} or the minimal local Hilbert space dimension of each party in a Bell test necessary for a given violation \cite{dimWitOr}. The basic underlying approach to all these tasks is the same: we consider a property (entanglement, nonlocality, dimensionality, etc.) of the system that we wish to falsify, derive an operational limitation on the set of all systems satisfying this property and then show that this limitation is violated in the experiment.

Effective numerical tools to derive the operational limitations of small systems \cite{npa,finiteNPO,finiteNPO2,inflation,DPS1,DPS2,DPS3} are available. These allow detection of global properties such as entanglement and nonlocality in three or even four-partite systems in a few minutes using a regular desktop computer. Unfortunately, the analysis of large systems presents two problems.

The first is that the number of parameters required to fully specify the operational behavior of a many-body system increases exponentially with its size. So do the resources (e.g.: number of experiments) needed to estimate all such parameters. That is, even prior to detection, one cannot  efficiently specify the state of the system in general. However, many natural quantum states admit an efficient \textit{tensor network} representation, see e.g. \cite{Werner,cirac,hastings,Vidal,peps,TNReview}, which has been exploited to develop tomographic protocols to characterize such states with a number of experiments that scales only polynomially with the system size \cite{tomo1,tomo2}. Assuming that the quantum states underlying our experiments are somehow typical, and can be represented by a tensor network, it is therefore possible to circumvent this problem. 

The second problem is that, even when the system can be efficiently represented, the computational resources required to detect the relevant global properties of the system also scale exponentially with the system size. Consequently, as experiments with quantum simulators and condensed matter systems progress, we get access to larger and larger systems whose non-classical properties cannot be detected with current theoretical tools.

In this work, we propose a general approach to solve this second problem. Our approach will rely heavily on the framework of tensor networks, and also, somewhat surprisingly, on a central concept from quantum foundations: the framework of Generalized Probabilistic Theories (GPTs) \cite{GPT1,GPT2,GPT3,GPT4,GPT5}. It will also require techniques from convex optimization theory \cite{lp,sdp}. The result, \textit{connector theory}, will allow us to detect global properties of many-body systems via algorithms whose time and memory complexity scales \textit{linearly} with the system size. This lets us access systems made of hundreds of sites.

The key insight underlying connector theory is that one can construct local transformations to \textit{coarse-grain} the many-body system to an effective small system, say with 2 or 3 sites, while preserving the global property of interest. Subsequently, if one can detect that the resulting small system has the desired property, then so does the original system.

Our inspiration comes from renormalisation approaches and, in particular, coarse-graining techniques that have proved very effective in e.g. diagonalizing large quantum-many body Hamiltonians in condensed matter physics where one is often interested only in the low energy subspace. The many-body Hamiltonian can be coarse-grained to a few-site Hamiltonian---which can be exactly diagonalized---while preserving its low energy subspace. This strategy has led to the invention of ground breaking simulation algorithms for large condensed matter systems.

Before presenting all the details of the method, it is useful to illustrate the main idea with an example. Suppose you are given a quantum state $\rho$ of $m$ particles and your task is to certify that the state is entangled. An $m$-body quantum state is separable if it can be expressed as
\be
\rho=\sum_i p_i\rho_i^1\otimes...\otimes\rho_i^m,
\label{sepState}
\ee
\noindent where $p_i\geq 0$, $\sum_i p_i=1$, and $\rho_i^j$ are normalized quantum states. The state $\rho$ is entangled if it does not admit such a decomposition. Now, assume you have a linear transformation mapping two systems into one, 
\begin{equation}
T : (\mathbb{C} \otimes \mathbb{C} ) \rightarrow \mathbb{C},  
\end{equation}
and such that product states are transformed into valid quantum states, that is $T(\rho\otimes\sigma)\geq 0$ for all $\rho$ and $\sigma$. Note that we don't require the map to be physical, that is, it may produce non-positive states when applied to an initial (entangled) state. Clearly the application of this map to a separable $m$-body state results into a separable $(m-1)$-body state. By repeatedly applying maps of this form to the initial state, it is possible to reach a size in which standard entanglement detection methods, including state positivity, can be tested. If any of these methods fails, we can certify that the initial $m$-body state was entangled. However, to apply this idea in practice, many aspects need to be sorted out. For instance, one needs to find a way of applying the maps $T$ without having to deal with the complete $m$-body quantum state and provide the tools to construct them. These and other issues are presented in what follows and constitute the main technical results of this work.

The article is organized as follows. In Section \ref{background}, we begin by reviewing the three essential ingredients of connector theory: tensor networks (as efficient representation of quantum many-body states), convex optimization and the formalism of GPTs. This section also introduces the graphical notation for tensor networks that is convenient to explain the basics of connector theory, and is used in the rest of the paper. In Section \ref{Bell}, we explain the use of connector theory for detecting Bell nonlocality as a case study. In Sections \ref{supraQ}, \ref{secEnt}, we describe how to apply the formalism of connector theory to detect supraquantum nonlocality and entanglement in quantum many-body systems. Finally, in section \ref{conclusion}, we present our conclusions.

\section{Background}
\label{background}

The main objective of this first section is to review the main ingredients used in our construction: tensor networks, techniques from convex optimization theory and generalized probabilistic theories (GPT).

\subsection{Tensor networks}
We start by providing a broad introduction to the formalism of tensor networks as it appears in quantum information theory and condensed matter physics for efficiently representing quantum many-body states. For a review, see \cite{TNReview}. 

\begin{figure}
  \centering
  \includegraphics[width=6.5cm]{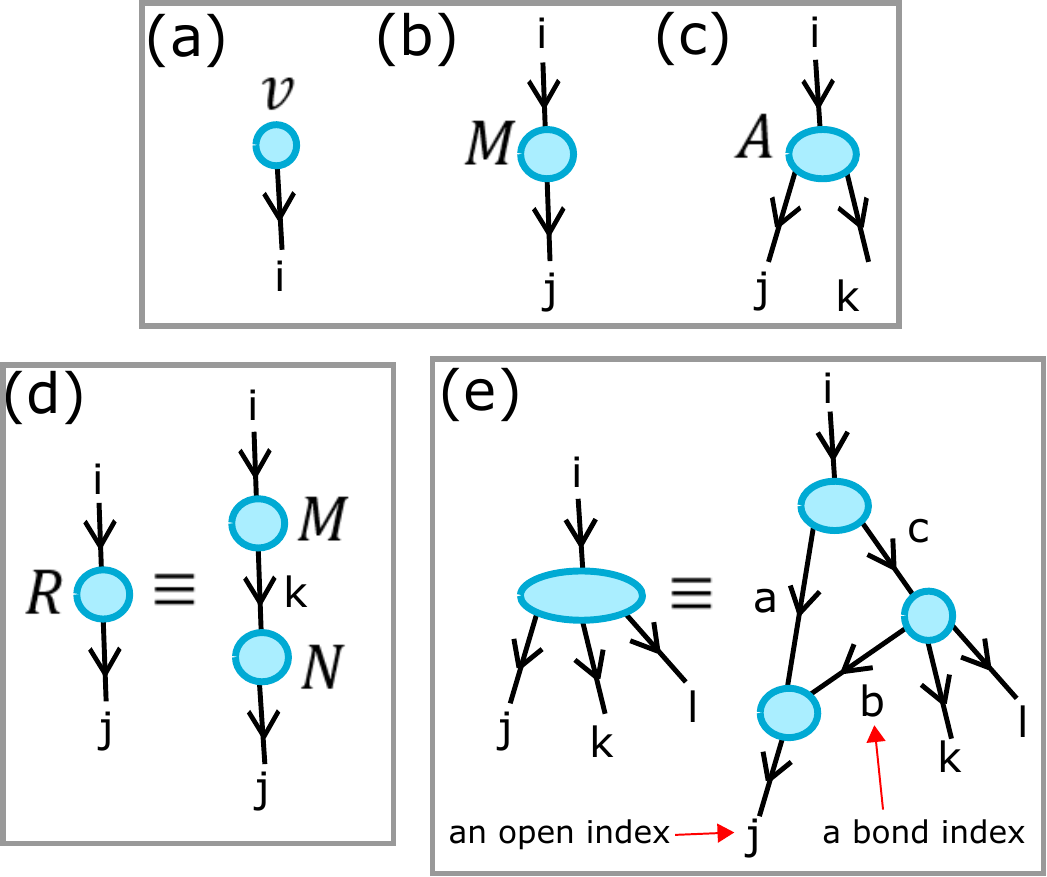}
  \caption{\textbf{Graphical representation of tensors and tensor operations.} (a) A vector $\ket{v} \equiv \sum_i v_i \ket{i}$. (b) A matrix $M \equiv \sum_{ij}M_{ij}\ket{j}\bra{i}$. (c) A 3-index tensor $T = \sum_{ijk} T_{ijk} \ket{j}\ket{k}\bra{i}$. (d) Matrix multiplication, \eref{eq:matrixmult}. (e) Example of a more general tensor contraction, \eref{eq:tensorcontract}. Two types of indices are distinguished in a contraction: \textit{open indices} that emanate only from one tensor and are left uncontracted, and \textit{bond indices} that connect two tensors and gets summed over.}
  \label{fig:tensors}
\end{figure}

Broadly speaking, a tensor network is a set of \textit{tensors} that are interconnected or contracted according to a given network. By a tensor we simply mean a multi-dimensional array of complex coefficients---an object that generalizes the notion of vectors and matrices.
More precisely, a tensor $T^{i_1i_2...i_m}_{j_1j_2...j_n}$ is a linear map from the tensor product of a set of input vector spaces to the tensor product of some output vector spaces,
\begin{equation}
T : (\mathbb{V}_{1} \otimes \mathbb{V}_{2} \otimes \ldots \mathbb{V}_{m}) \rightarrow (\mathbb{W}_{1} \otimes \mathbb{W}_{2} \otimes \ldots \mathbb{W}_{n})
\end{equation}
The indices $i_1,i_2,\ldots,i_m$ label an orthonormal basis in the input spaces $\mathbb{V}_{1}, \mathbb{V}_{2},\ldots, \mathbb{V}_{m}$ respectively, whereas the indices $j_1,j_2,\ldots,j_n$ label an orthonormal basis in the output spaces $\mathbb{W}_{1}, \mathbb{W}_{2},\ldots, \mathbb{W}_{n}$ respectively. For convenience, we represent tensors graphically as illustrated in \fref{fig:tensors}. A tensor is depicted by a shape and its indices are depicted by directed lines emanating from the shape. Input and output indices, which we later need to distinguish, are indicated by attaching incoming and outgoing arrows to the corresponding lines.

One can obtain a new tensor by \textit{contracting} or multiplying together a set of tensors. Tensor contraction generalizes the notion of matrix multiplication. Two matrices $M$ and $N$ can be multiplied to obtain a new matrix $R \equiv MN$. In tensor notation we write,
\begin{equation}\label{eq:matrixmult}
R^{i}_j = \sum_k M^i_k N^k_j.
\end{equation}
We graphically depict this by connecting an output index of matrix $M$ with an input index of matrix $N$, as shown in \fref{fig:tensors}(d). Of course, the dimension of the output index of $M$ must equal the dimension of the input index of $N$. The \textit{dimension} of an index is the number of values the index runs over. For example, if $R$ in \eref{eq:matrixmult} is a $2 \times 3$ matrix then the dimension of indices $i$ and $j$ is equal to $2$ and $3$ respectively. 

A more general tensor contraction is illustrated in \fref{fig:tensors}(e), where three tensors $A,B$ and $C$ are contracted to obtain a 4-index tensor $T$,
\begin{equation}\label{eq:tensorcontract}
T^{i}_{jkl} = \sum_{abc} A^i_{ac} B^{ab}_jC^c_{bkl}.
\end{equation}
In a contraction, the indices that are left uncontracted are called \textit{open indices} e.g. $i,j,k$ and $l$. On the other hand, indices that are summed over are called \textit{bond indices}, e.g. $a,b$ and $c$.

The generic pure quantum state $\ket{\Psi}$ of a large many-body system e.g. a lattice of qubits, requires specifying $2^N$ probability amplitudes, where $N$ is the number of qubits:
\begin{equation}\label{eq:manybodystate}
\ket{\Psi} \equiv \sum_{i_1,i_2,\ldots,i_N} \Psi_{i_1,i_2,\ldots,i_N} \ket{i_1} \otimes \ket{i_2} \otimes \ldots \otimes \ket{i_N}.
\end{equation}
Even more parameters, of the order of $4^N$, are needed if the state is mixed. However, often interesting states such as ground states or thermal states of local Hamiltonians contain a limited amount of correlations. This can be exploited to efficiently represent them by decomposing the large quantum many-body wavefunction encoded in the exponentially $N$-index tensor $\Psi_{i_1,i_2,\ldots,i_N}$ into a product of small tensors, namely, as a tensor network.

\begin{figure}
  \centering
  \includegraphics[width=6cm]{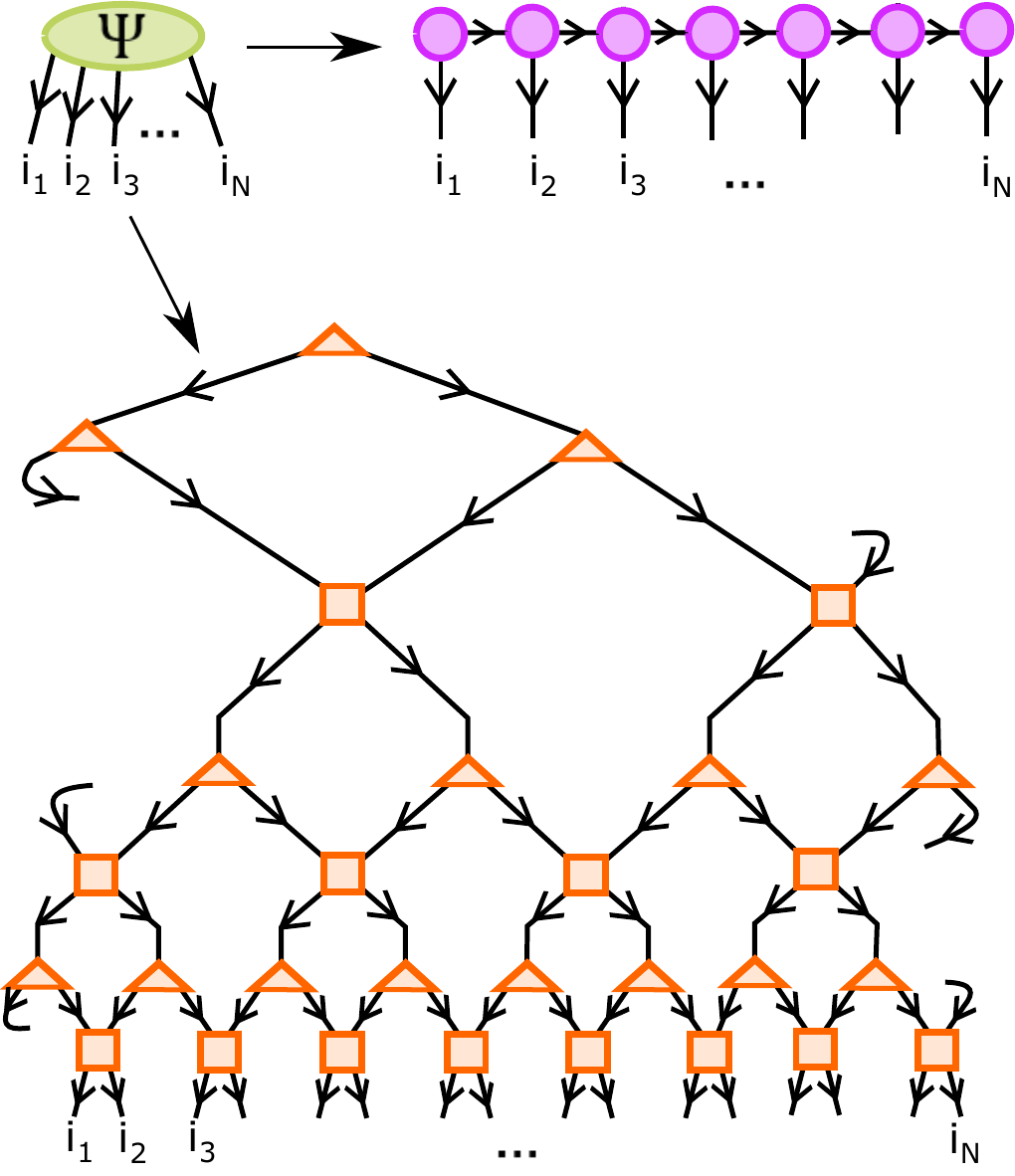}
  \caption{\textbf{Examples of popular tensor network decompositions of quantum many-body states.} (a) A $N$-index tensor $\Psi$ that e.g. stores the probability amplitudes of a quantum many-body state in \eref{eq:manybodystate}. (b) A matrix product state (MPS) decomposition of tensor $\Psi$. It corresponds to a tensor network composed from tensors with at most 3 indices. Tensor $\Psi$ is recovered by contracting together the tensor network. (c) A different and more elaborate tensor network decomposition of $\Psi$, called the MERA, which is composed of 3-index tensors (triangles) and 4-index tensors (squares). (Once again, tensor $\Psi$ is recovered by contracting together the tensor network.)}
  \label{fig:TN}
\end{figure}

\fref{fig:TN} illustrates two popular tensor network decompositions---matrix product states (MPSs) \cite{Werner,cirac} and the multi-scale entanglement renormalization ansatz (MERA) \cite{Vidal}---that have been used to efficiently represent ground state of local Hamiltonians acting on a one-dimensional quantum lattice. Later, we will propose the use of connector tensor networks that are structurally similar to these decompositions.

Here, the sites of the lattice correspond to the open indices of tensor networks, while the bond indices carry the entanglement and correlations in the state. The dimensions of the bond indices generally indicate the amount of entanglement and correlations in the state: a larger bond dimension generally corresponds to larger entanglement and correlations. By contracting together all the tensors in a tensor network one recovers the probability amplitudes in \eref{eq:manybodystate}. The choice of the network pattern of the tensor network decomposition of a given state is dictated by the specific structure of entanglement in the state.

In a condensed-matter context, one typically uses these tensor networks as an ansatz for the unknown ground state (or a low energy subspace) of a given Hamiltonian and determines the tensors numerically by means of, say, a variational energy minimization. The maximum bond dimension in a tensor network ansatz determines both the cost of the numerical optimization and the accuracy of the approximation. Here, we propose a novel application of tensor networks in the context of certification of relevant quantum properties, for instance, as witnesses for entanglement and non-locality. Moreover, and as described below, these tensor networks can be understood as measurements in a general probabilistic theory, thus extending the formalism of tensor networks beyond quantum theory.

\subsection{Convex optimization theory}
Let $X$ be a vector space, and let ${\cal X}\subset X$ be a convex subset thereof. The goal of convex optimization is to solve problems of the form

\begin{align}
&\min f(\bar{x})\nonumber\\
\mbox{such that }& \bar{x}\in {\cal X},
\end{align}
\noindent where $f$ is a convex function, i.e., $f(p\bar{x}_1+(1-p)\bar{x}_2)\leq pf(\bar{x}_1)+(1-p)f(\bar{x}_2)$, for $\bar{x}_1,\bar{x}_2\in{\cal X}$, $0\leq p\leq 1$. Any vector of variables $\bar{x}\in X$ satisfying $\bar{x}\in {\cal X}$ is said to be a \emph{feasible point}.

\emph{Linear programming} (LP) \cite{lp} is a branch of convex optimization where ${\cal X}$ is a polytope (a convex set defined by a finite number of linear inequalities) and $f$ is a linear function of the variables $\bar{x}\in \R^n$ of the optimization problem. Linear programming is thus concerned with optimization problems of the sort:

\begin{align}
&\min \bar{c}\cdot\bar{x}\nonumber\\
\mbox{such that }& A\bar{x}\geq \bar{b}.
\label{standLP}
\end{align}
Here the $m\times n$ matrix $A$, $\bar{b}\in\R^n$ and $\bar{c}\in \R^m$ are the inputs of the problem. For any pair of vectors $\bar{y}, \bar{z}$ of identical size, the notation $\bar{y}\geq \bar{z}$ indicates that $y_i-z_i\geq 0$ for all $i$. As we will see, linear programming is an instrumental tool for nonlocality detection. 

In order to deal with entanglement and quantum nonlocality, we use a more sophisticated tool, namely \emph{semidefinite programming} (SDP) \cite{sdp}. A semidefinite program is an optimization problem of the form:

\begin{align}
&\min \bar{c}\cdot\bar{x}\nonumber\\
\mbox{such that }& F_0+\sum_{i=1}^nx_iF_i\geq 0.
\label{standSDP}
\end{align}
This time the $m\times m$ matrices $F_0, \{F_i\}$ and the vector $\bar{c}$ constitute the problem input. Beware the change in notation: if $A$ is a square matrix, then $A\geq 0$ is used to denote that $A$ is positive semidefinite, i.e., it is self-adjoint and all its eigenvalues are non-negative.

There exist free solvers available to solve both linear and semidefinite programs. These solvers exploit convex optimization theory to provide, not only an approximate solution of the problem, but also rigorous bounds on how this figure differs from the exact value. For linear programs of any size, we recommend the MATLAB solver Gurobi \cite{gurobi}; the packages Sedumi \cite{sedumi} and Mosek \cite{mosek} are appropriate, respectively, to solve small and large instances of semidefinite programs. We recommend not to work with these solvers directly, but through general optimization MATLAB packages, such as YALMIP \cite{yalmip} or CVX \cite{cvx,cvx2}. The advantage of using either of these packages is that the user does not need to write the programs in the standard form (\ref{standLP}), (\ref{standSDP}): it is enough to indicate what linear or semidefinite constraints the variables $\bar{x}$ of the problem must be subjected to.

Unless otherwise specified, in all our numerical computations we make use of YALMIP \cite{yalmip} in combination with Gurobi \cite{gurobi} (for LPs) or Mosek \cite{mosek} (for SDPs).

\subsection{Generalized probabilistic theories}

The formalism of GPTs was conceived to reason about physical theories beyond quantum physics. In a sense, it conveys an operational description of what one can do within a physical theory, but without a correspondence principle to relate the mathematical formalism of the theory to the instruments of an experimental workshop. Viewed as a GPT, quantum physics is a theory where each system is labeled by a natural number $D$ (the dimension). Normalized (subnormalized) states of a system of dimension $D$ are described by $D\times D$ complex positive semidefinite matrices with trace (smaller than or equal to) $1$; measurements are defined by Positive Operator Valued Measures (POVMs); and transformations, by completely positive trace-preserving maps. Also, states of a bipartite system of dimensions $D,D'$ are in one-to-one correspondence with the states of a system of dimension $DD'$.

More generally, a GPT is specified by a list of possible system types, together with composition rules specifying which system type describes the combination of several other types. In a GPT the state of a given system of type $S$ is identified with a vector \footnote{In quantum theory, one can construct this vector by writing density matrices in vector form.} $\bar{v}$, living in a space $\H_S$. The set of possible states of $S$ corresponds to a convex set ${\cal C}_S\subset \H_S$. For every system $S$ we assume the existence of a vector $\bar{e}_S\in\H_S$, the \emph{unit effect}, whose scalar product with any state returns the \emph{norm} of the state, or the probability that the state was successfully prepared. It follows that, for all $\bar{v}\in{\cal C}_S$, $\bar{e}_S\cdot\bar{v}\leq 1$. Moreover, $\bar{v}\in{\cal C}_S$ is a deterministic preparation iff $\bar{e}_S\cdot\bar{v}=1$. Sometimes, for simplicity, we use the notation $E(\bar{v})=\bar{e}_S\cdot \bar{v}$.

In the following we will only consider GPTs which satisfy local tomography \cite{GPT1}. In our language, this implies that $\H_{S\otimes S'}=\H_S\otimes\H_{S'}$, where $S\otimes S'$ denotes the composition of systems $S,S'$. To recover the marginal state $\bar{v}_S$ of system $S$ from the joint state $V_{SS'}$ of systems $S, S'$, we apply the unit effect over the space $\H_{S'}$, i.e., $\bar{v}_S=\id_S\otimes\bar{e}_{S'}\cdot \bar{v}_{SS'}$.

Any (non-deterministic) transformation of system $S$ into another system of type $S'$ corresponds to a linear map $W:\H_S\to\H_{S'}$ with the property that, for any system type $T$,

\begin{align}
\bar{v}_{S\otimes T}\in {\cal C}_{S\otimes T}\Rightarrow &(W\otimes \id_T)\cdot\bar{v}_{S\otimes T}\in {\cal C}_{S'\otimes T},\nonumber\\
&E\left(W\bar{v}_{S}\right)\leq E\left( \bar{v}_{S}\right).
\label{NRH}
\end{align}
\noindent When the output system $S'$ has dimension $1$, the transformation corresponds to a vector $\bar{w}\in \H_S$, and it physically represents an \emph{effect}. The probability that the event signified by $\bar{w}$ occurs is then given by $\bar{w}\cdot \bar{v}_S$.

Effects must not be confused with \emph{witnesses}. A normalized witness is a vector $\bar{w}\in \H_S$ with the property $0\leq \bar{w}\cdot \bar{v}\leq 1$ for all $\bar{v}\in {\cal C}_S$. An effect has, in addition, the property that $(\bar{w}^T\otimes\id_T)\cdot\bar{v}_{ST}$ is a state in ${\cal C}_T$ if $\bar{v}_{ST}\in {\cal C}_{ST}$. While all effects are normalized witnesses, (in general GPTs) not all normalized witnesses are effects.

\section{A case study: Bell nonlocality}
\label{Bell}
Once the main building blocks of the construction have been presented, in what follows we illustrate how to use connector theory to detect relevant properties of large systems. We do so by means of different relevant examples, starting by the detection of Bell nonlocality. We explain the connector formalism in this scenario, characterize connectors for detecting Bell non-locality and techniques for optimizing them. After non-locality in the following sections, we also show how to use connectors in the context of supra-quantum nonlocality and entanglement detection.

\begin{figure}
  \centering
  \includegraphics[width=4 cm]{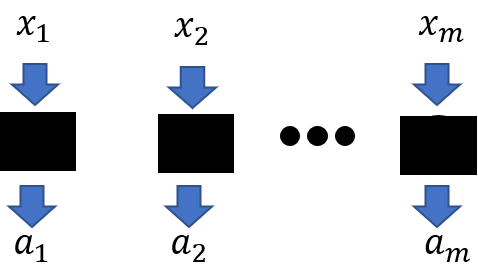}
  \caption{\textbf{Bell test.} A Bell test consists of $m$ distant parties performing different measurements on their systems. The choice of measurements by party $i$ is labelled by $x_i$ and the corresponding output by $a_i$. The different systems are seen as back boxes, each  producing the classical output $a_i$ after receiving the classical input $x_i$. The whole scenario is described by the conditional probability distribution $P(a_1,...,a_m|x_1,...,x_m)$.}
  \label{boxPic}
\end{figure}

\begin{figure}
  \centering
  \includegraphics[width=7cm]{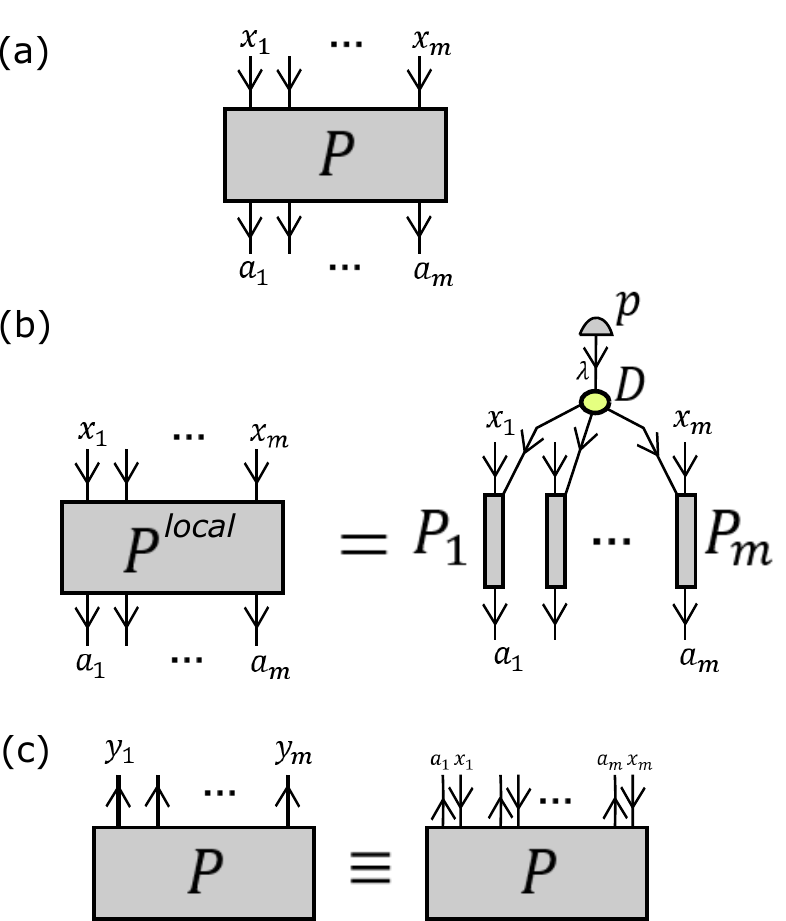}
  \caption{\textbf{Boxes as tensors.} (a) A no-signalling box as a tensor $P^{x_1x_2\ldots x_m}_{a_1a_2\ldots a_m}$. (b) A Bell-local box $P^{local}$ decomposes as shown in terms of 3-index tensors $P_1,P_2,\ldots,P_m$, each of which is a conditional probability distribution. Here $D$ is a $m-$index \textit{identity tensor} (also called a \textit{copy tensor}), namely, a tensor whose only non-zero components are $D_{ii\ldots i}=1$, and $p_\lambda$ is a vector of probabilities $p(\lambda)$. (d) For convenience, we will combine each pair $(a_k,x_k)$ of input and output indices of a box into a single outgoing index $y_k \equiv (a_k,x_k)$.}
  \label{fig:boxes}
\end{figure}

Consider an $m$-partite Bell scenario, where each party interacts with a black box, to which it can input a symbol $x$ and then obtain an output $a$, see \fref{boxPic}. The operational description of this box is given by the probabilities $P(a_1,...,a_m|x_1,...,x_m)$. We will represent these probabilities also as a \textit{tensor} with $m$ incoming and $m$ outgoing indices, namely,
\be
P^{x_1...x_m}_{a_1...a_m} \equiv P(a_1,...,a_m|x_1,...,x_m),
\ee
and graphically represent this tensor as shown in \fref{fig:boxes}(a). We assume that this box is \emph{non-signalling}, i.e.,  for any $k\in\{1,...,m\}$, the marginal probability distribution
\be
\sum_{a_k}P^{x_1...x_m}_{a_1...a_m}
\ee
\noindent does not depend on $x_k$. 

We are asked whether box $P$ is \textit{Bell local} i.e., whether the correlations $P^{x_1...x_m}_{a_1...a_m}$ can be expressed as
\begin{align}
P^{x_1...x_m}_{a_1...a_m} =\sum_{\lambda} p_{\lambda} {(P_1)}^{x_1,\lambda}_{a_1}...{(P_m)}^{x_m,\lambda}_{a_m},
\label{localBox}
\end{align}
where $p(\lambda), {(P_j)}^{x_j,\lambda}_{a_j} \equiv P_j(a_j|x_j,\lambda)$ are arbitrary probability distributions, see \fref{fig:boxes}(c). We also refer to Bell local boxes as \textit{classical boxes}.

To help us answer this question, we introduce a GPT, that we call \textbf{LOC-world}. Intuitively, each system in \textbf{LOC-world} corresponds to a multipartite box, with a number of possible inputs and outputs. A general system is thus labelled by a vector of natural numbers of the form $[O_1,...,O_m,I_1,...,I_m]$, which denotes that the $k^{th}$ party's box has $I_k$ inputs and $O_k$ possible outputs. In \textbf{LOC-world}, the set of states of any system of type $[O_1,...,O_m,I_1,...,I_m]$ corresponds to the set of unnormalized probabilities  $P^{x_1...x_m}_{a_1...a_m}$ of the form (\ref{localBox}). We define the norm of a state $P$ in \textbf{LOC-world} as $E(P)$,
\be
E(P)=\sum_{a_1,...,a_m}P^{0...0}_{a_1...a_m}. 
\ee
Valid transformations $W$ in \textbf{LOC-world} correspond to linear maps which, acting on part of a classical box $P$, return a classical box $P'=(W\otimes \id)P$. To be interpreted as non-deterministic transformations, such maps must satisfy the condition $E(P')\leq E(P)$. We call such transformations \emph{connectors} (in \textbf{LOC-world}). 

\begin{figure}
  \centering
  \includegraphics[width=\columnwidth]{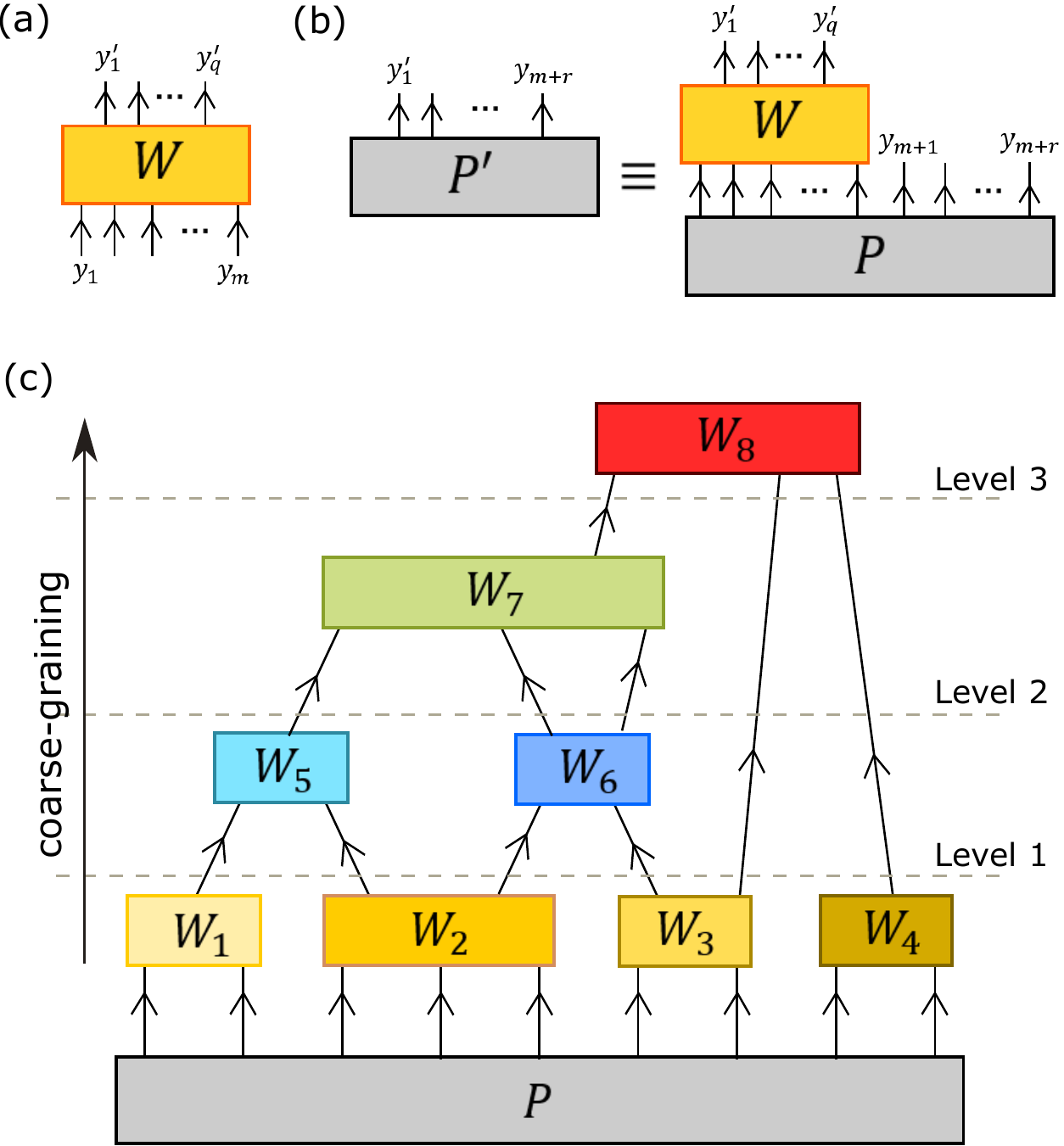}
  \caption{\textbf{Connectors.} (a) Graphical representation of a $m\to q$ connector $C$. (b) Contraction of $C$ with an $(m+r)$-partite non-signalling box $P$ results in a $(q+r)$-partite non-signalling box $P'$.(c) Connectors can be used to coarsegrain a box to a manageable size. The coarse-grained box at any `level' (indicated by a dashed line) is obtained by contracting the original box $P$ with all the connectors up to that level. Here, at Level 3 we obtain a 3-party box, whose nonlocality can be probed exactly with a known 3-party witness $W_8$ (red).}
  \label{fig:connector}
\end{figure}

A general $m\to q$ connector takes as input a system of type $[O_1,...,O_m,I_1,...,I_m]$ and returns as output a system of type $[O'_1,...,O'_q,I'_1,...,I'_q]$. Any $m\to q$ connector can be interpreted as a tensor of the form $W_{y_1...y_m}^{y'_1...y'_q}$ (notice that all indices are doubled indices). This tensor is graphically represented as shown in \fref{fig:connector}(a). Acting with $W$ on a $(m+r)$-partite box $P$ results in a $(q+r)$-partite box $P$. The action of $W$ over the first $m$ systems of $P$ can be expressed as the contraction of the first $m$ outgoing legs of $P$ with the incoming legs of $C$:
\begin{align}
P'_{y'_1,...,y'_q,y_{m+1},...,y_{m+r}} \equiv \sum_{y_1,...,y_m}W_{y_1...y_m}^{y'_1...y_q'}P_{y_1...y_{m+r}},
\end{align}
depicted in \fref{fig:connector}(b).

Now, suppose that, given a non-signalling box $P$ for 9 parties, we applied to it connectors $W^1,W^2,...,W^8$ as depicted in \fref{fig:connector}(c). Consider the case in which the output system of the top most connector $W^8$ is of type $[1,1]$, resulting on a joint probability distribution for a one-input and one-output situation, that is, a probability $0\leq p\leq 1$. Then the action of all the connectors $W^1,W^2,...,W^8$ can be interpreted as a measurement $W$ in \textbf{LOC-world}. Clearly, if $W(P)<0$ or $W(P)>1$, it follows that $P$ did not belong to the class of states of \textbf{LOC-world}. In other words: $P$ is nonlocal. 

This observation is the basis of connector theory. Namely, any \textit{tensor network of connectors} that has

\begin{enumerate}
\item
no outgoing arrows; and
\item
no cycles, 
\end{enumerate}
\noindent defines a normalized Bell inequality \cite{almostNRH}, i.e., a linear functional with the property that $0\leq W(P)\leq N(P)$ for all classical boxes $P$. It is easy to see that any attempt at enlarging the set of connectors with extra tensors will result on the loss of this property. 

Note that in principle there is a more general strategy to detect that our original tensor does not belong to the set of states of the theory. Namely, use connectors to coarse-grain the system and then apply a witness to prove the non-physicality of the resulting coarse-grained network. For instance, in Figure \ref{fig:connector} (c) we would replace the connector $W_8$ by an arbitrary normalized witness. Since the set of normalized witnesses contains the set of effects in a GPT, this method should allow us to detect more instances of Bell nonlocality. However, in \textbf{LOC-world}, as well as in the other two GPTs we define in the following for supra-quantum and entanglement detection, normalized witnesses happen to be connectors as well, so it is enough when we consider connector tensor networks.

\subsection{Characterizing connectors in LOC-world}

Due to the structure (\ref{localBox}) of the set of classical boxes, \textbf{LOC-world} has the convenient property that any linear map fulfilling condition (\ref{NRH}) with $T=\emptyset$ constitutes a valid transformation. That is, if a given map in \textbf{LOC-world} is valid when acting on a system, it is also a valid map when acting on parts of a larger system. Let us see why. 

Any box of the form $P^x_a$ can be expressed as 

\be
P^x_a=\sum_{\bar{a}}p_{\bar{a}} P^{x,\bar{a}}_a,
\ee
\noindent where $p_{\bar{a}}$ is a probability distribution over $\{1,...,d\}^n$ and $P^{x,\bar{a}}_a=\delta_a^{a_x}$ are deterministic boxes. Absorbing $p_{\bar{a}}$ in the definition of the hidden variable $\lambda$ in eq. (\ref{localBox}), we have that an $m$-partite box is classical iff it can be expressed as a convex combination of $m$-partite deterministic boxes of the form $P_{a_1,...,a_m}^{x_1,...,x_m}=\prod_{k=1}^mP_{a_k}^{x_k,\bar{a}^k}$. Each of these deterministic boxes is an \emph{extreme point}, namely, a point that cannot be decomposed as a convex decomposition of other points within the set of classical boxes.

Given an $m\to q$ connector $\Omega$, deciding whether it satisfies (\ref{NRH}) amounts to verifying that $\Omega\otimes \id_T$ maps deterministic boxes to classical boxes. The general result then follows by applying convexity. Now, suppose that $\Omega$ satisfies (\ref{NRH}) with $T=\emptyset$, consider an arbitrary $m+r$ partite deterministic box $P\equiv P^1P^2...P^{m+r}$ and let $\Omega$ act over the first $m$ systems (equivalently, take $T=\{m+1,...,m+r\}$). The result will be the (in general, unnormalized) box $P'=QP^{m+1}...P^{m+r}$, with $Q=\Omega(P^{1}...P^{m})$. By hypothesis, $Q$ is classical, and therefore so is $P'$. We thus conclude that $\Omega\otimes \id_T$ will map classical boxes to classical boxes for all $T$.

Note that this property does not hold in quantum mechanics. Indeed, there exist positive linear maps $\Omega$, like the transposition map $\Omega(\rho)=\rho^T$, which satisfy $\Omega(\rho)\geq 0$ for all $\rho\geq 0$ (i.e., they are positive), despite the fact that $\Omega\otimes \id$ is not positive \cite{jami,choi}. In other words, in the case of local correlations, it is impossible to find maps that are analogue to the positive but not completely positive maps for quantum states.

We have thus reduced the problem of characterizing connectors in \textbf{LOC-world} to the problem of identifying those transformations which map extreme classical boxes to classical boxes. Since classical boxes form a convex set with a finite number of extreme points, it follows that characterizing or conducting linear optimizations over connectors in \textbf{LOC-world} can be cast as a \textit{linear program} (LP). See Appendix \ref{appLoc} for a detailed description of the LPs, together with some tips to reduce their time and memory complexity.

\begin{figure}
  \centering
  \includegraphics[width=3 cm]{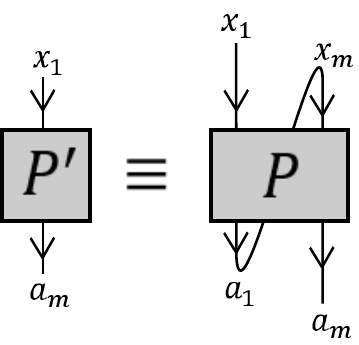}
  \caption{\textbf{Example of a wiring.} A 2-party box $P$ is mapped into a single party box $P'$ by using as input for the second system the output generated by the first.}
  \label{fig:wiring}
\end{figure}

Whats' the form of connectors in \textbf{LOC-world}? Some of them correspond to \emph{wirings} \cite{wirings}, namely, transformations that correspond to feeding the outputs of some parties to the inputs of some other parties. To fix ideas, consider connectors from $(2,2,2,2)$ systems to $(2,2)$ systems, with input indices $(a_1,x_1),(a_2,x_2)$ and output indices $(b,y)$. If we denote by $P,P'$ the input and output boxes, then a possible wiring would be given by 
\be
{P'}_b^y=\sum_a P_{a,b}^{y,a}=\sum_{a,a',x,x'}W^{b,y}_{a,a',x,x'}P_{a,a'}^{x,x'},
\ee
\noindent with $W^{b,y}_{a,a',x,x'}=\delta_{y,x}\delta_{b,a'}\delta_{a,x'}$. This is just the result of inputting $y$ on the first part of $P$, reading the result $a$ and using it as an input in the second part of $P$. The final outcome of such an effective box is the output $b$ produced by the second part of $P$, see \fref{fig:wiring}. Although wirings map non-signalling boxes to non-signalling boxes---and therefore they are examples of connectors---the contraction of any network of wirings with a non-signalling box always results in a non-negative number. This means that wirings, by themselves, cannot be used to detect Bell nonlocality. Fortunately, there exist more general connectors in \textbf{LOC-world}, as shown next.

\subsection{Connectors built from Bell inequalities}
Consider the following $2$-to-$1$ connector defined by the relations:
\begin{align}
&C_{axa'x'}^{0,0}={\cal C}_{axa'x'}, C_{a,x,a',x'}^{1,0}=(E-{\cal C})_{axa'x'}\nonumber\\
&C_{a,x,a',x'}^{0,1}={\cal C}'_{axa'x'},C_{a,x,a',x'}^{1,1}=(E-{\cal C}')_{axa'x'},
\label{connC}
\end{align}
\noindent where ${\cal C}, {\cal C}'$ correspond to normalized forms of the Clauser-Horne-Shimony-Holt (CHSH) Bell inequality, i.e.,

\begin{align}
{\cal C}_{axa'x'}\equiv&-\delta_a^0\delta_{a'}^0(\delta_x^0\delta_{x'}^0+\delta_x^1\delta_{x'}^0+\delta_x^0\delta_{x'}^1-\delta_x^1\delta_{x'}^1)\\\nonumber
&+\delta_x^0\delta_{x'}^0(\delta_a^0+\delta_{a'}^0),\nonumber\\
{\cal C}'_{axa'x'}\equiv&-\delta_a^0\delta_{a'}^0(-\delta_x^0\delta_{x'}^0+\delta_x^1\delta_{x'}^0+\delta_x^0\delta_{x'}^1+\delta_x^1\delta_{x'}^1)\nonumber\\
&+\delta_x^1\delta_{x'}^1(\delta_a^0+\delta_{a'}^0).
\label{CHSHs}
\end{align}
\noindent It can be verified that $0\leq {\cal C}(P),{\cal C}'(P)\leq 1$ for any bipartite classical probability distribution $P$ with two inputs and two outputs. It follows that, for any classical box $P$, the new box $P'=C(P)$ will be such that ${P'}_b^y\geq 0$, for $b,y=0,1$ and $\sum_{b}{P'}_b^y=E(P)$. This connector corresponds to a deterministic transformation in \textbf{LOC-world}.

How would we implement transformation $C$ in practice? $C$ is neither a wiring nor a convex combination thereof: this follows from the fact that, applied over any box violating the CHSH inequality, it will return a `box' with negative probabilities. Now, suppose that the input box $P$ is indeed classical. This implies that, hidden within $P$, there exist variables $(a_0,a_1,a'_0,a'_1)\in\{0,1\}^4$ which will determine the outcomes of the box: if we input $x,y\in\{0,1\}$, we will obtain the outputs $a_x,a'_y$. The values of $(a_0,a_1,a'_0,a'_1)$ can change every time we initialize the box; we assume that they are distributed according to a measure $\mu(a_0,a_1,a'_0,a'_1)$. 

Now, consider the functions:
\begin{align}
&f(a_0,a_1,a'_0,a'_1)=-\bar{a}_0\bar{a}'_0-\bar{a}_1\bar{a}'_0-\bar{a}_0\bar{a}'_1+\bar{a}_1\bar{a}'_1+\bar{a}_0+\bar{a}'_0,\nonumber\\
&g(a_0,a_1,a'_0,a'_1)=\bar{a}_0\bar{a}'_0-\bar{a}_1\bar{a}'_0-\bar{a}_0\bar{a}'_1-\bar{a}_1\bar{a}'_1+\bar{a}_1+\bar{a}'_1,
\end{align}
\noindent where $\bar{c}\equiv 1-c$.
It is easy to check that $f(a_0,a_1,a'_0,a'_1),g(a_0,a_1,a'_0,a'_1)\in\{0,1\}$, for all $a_0,a_1,a'_0,a'_1\in\{0,1\}$. Moreover, 
\begin{align}
\sum_{a_0,a_1,a'_0,a'_1}\mu(a_0,a_1,a'_0,a'_1)f(a_0,a_1,a'_0,a'_1)={\cal C}(P),\nonumber\\
\sum_{a_0,a_1,a'_0,a'_1}\mu(a_0,a_1,a'_0,a'_1)g(a_0,a_1,a'_0,a'_1)={\cal C}'(P).
\end{align}
To implement $C$ in the lab over a local distribution $P$, it suffices to set up a device inside the box that can read the values $a_0,a_1,a'_0,a'_1$. On input $y=0$ ($y=1$), the device would return $b=0$, if $f(a_0,a_1,a'_0,a'_1)=1$ ($g(a_0,a_1,a'_0,a'_1)=1$), and $b=1$, otherwise. Obviously, such an operation is just possible if $P$ is a classical box to begin with. Inside a quantum box, for instance, $a_0,a_1$ could correspond to the outcomes of non-commuting measurements. Therefore, we could not have simultaneous access to them and the above scheme would be unrealizable.
\begin{figure}
  \centering
  \includegraphics[width=\columnwidth]{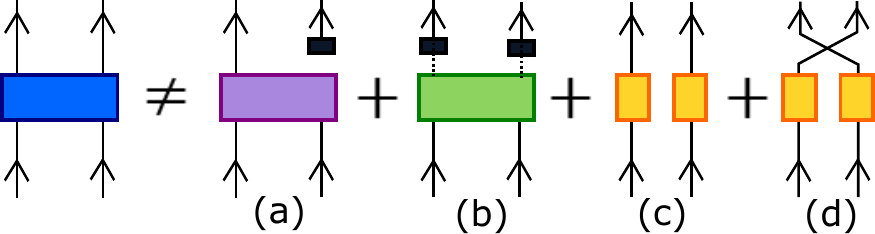}
  \caption{\textbf{Example of a non-trivial $2\to 2$ connector.} The $2\to 2$ connector identified in Appendix \ref{appDisEnt} cannot be decomposed as a combination of non-deterministic transformations of the form: a) A $2\to 1$ connector followed by a preparation; b) a (global) transformation of the bipartite box into a probability distribution, which we use as a local hidden variable model to build a new box; c) local mappings on both boxes; and d) local mappings followed by swapping the two parties.}  
  \label{disentPic}
\end{figure}

The fact that we used a Bell inequality to devise a $2\to 1$ connector is not coincidental. Actually, all $m\to 1$ connectors can be related to Bell inequalities. Let ${\cal B}$ be the set of $m$-partite classical boxes, and let ${\cal B}'$ be its dual, i.e., the set of linear functionals $U$ which map any box inside ${\cal B}$ to a non-negative number. Then all non-deterministic $m\to 1$ connectors $W$ in \textbf{LOC-world} are of the form

\begin{align}
&W_{c|z}\in {\cal B}',\sum_{c}W_{c|z}=W_0, E-W_0\in {\cal B}'.
\label{coarseB}
\end{align}
Now, it turns out that ${\cal B}'$ is in one-to-one correspondence with the set of Bell inequalities. Indeed, let $B(a_1,..,a_m,x_1,...,x_m), K\in\R$ be such that
\be
\sum_{\bar{a},\bar{x}}B^{\bar{a}}_{\bar{x}}P_{\bar{a}}^{\bar{x}} \geq K,
\ee
\noindent for all classical boxes $P_{a_1,...,a_m}^{x_1,...,x_m}$. Then the linear functional $U$ given by $U(P)\equiv\sum_{\bar{a},\bar{x}}B^{\bar{a}}_{\bar{x}}P_{\bar{a}}^{\bar{x}}-KE(P)$ satisfies $U(P)\geq 0$ for all classical boxes.

It is worth remarking that the notion of composing $m\to 1$ connectors to form new Bell inequalities is implicit in the work of Wu et al. \cite{marek}. There, the authors propose a scheme to generate a new $(m+1)$-partite new Bell inequality in the two-input/two-output Bell scenario, given two $m$-partite Bell inequalities. In our language, their scheme can be interpreted as a contraction between an $m\to 1$ connector and the Clauser-Horne-Shimony-Holt (CHSH) inequality \cite{chsh}.

To our knowledge, though, $m\to m'$ connectors have never been considered in Bell nonlocality, so we cannot relate them to past literature on the subject. A preliminary exploration of this class of transformations revealed rather intriguing objects. In this regard, in Appendix \ref{appDisEnt} we present a $2\to 2$ connector that does not admit a decomposition in terms of non-deterministic $1\to 1$ and $2\to 1$ connectors, see \fref{disentPic}. 
\begin{figure}
  \centering
  \includegraphics[width=\columnwidth]{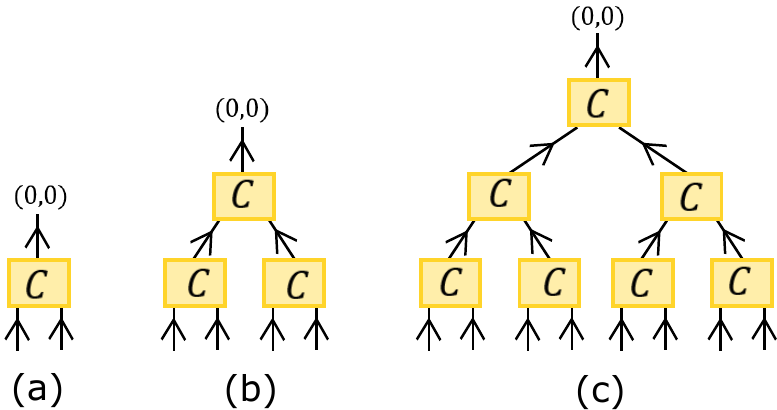}
  \caption{\textbf{CHSH trees.} Contracting many copies of connector (\ref{connC}) according to a tree network and inputting $(0,0)$ on the top connector, we obtain Bell inequalities for $n=2,4,8$ parties. The minimum non-signalling value of inequalities (a), (b), (c) are, respectively, -0.5, -1.5, -7.5.}
  \label{treePic}
\end{figure}

\subsection{Applications}
Now that we have non-trivial connectors, the next step is to contract them to generate new Bell inequalities. One possibility is to take $C$ in eq. (\ref{connC}) and contract multiple copies thereof in according to a tree network, see \fref{treePic}.

How useful are these new inequalities? As it turns out, computing the minimum value of an arbitrary Bell inequality under non-signalling distributions can be cast as a linear program. This allowed us to calculate, numerically, the corresponding maximal violations of each `CHSH tree', which seem to increase with the number of parties. Note that this value is a meaningful quantity, since all CHSH trees are normalized by construction.

However, the ultimate goal of Bell nonlocality detection is not to devise arbitrary Bell inequalities, but to detect the non-classicality of specific experimental systems. In this context, we now discuss two applications. First, detection of non-locality in  a experimental setup where the actual preparation of the underlying quantum state is known. And second, detection of non-locality for more general boxes.

\subsubsection{Nonlocality detection in finitely correlated states}\label{ssec:FCS}
\begin{figure}
  \centering
  \includegraphics[width=\columnwidth]{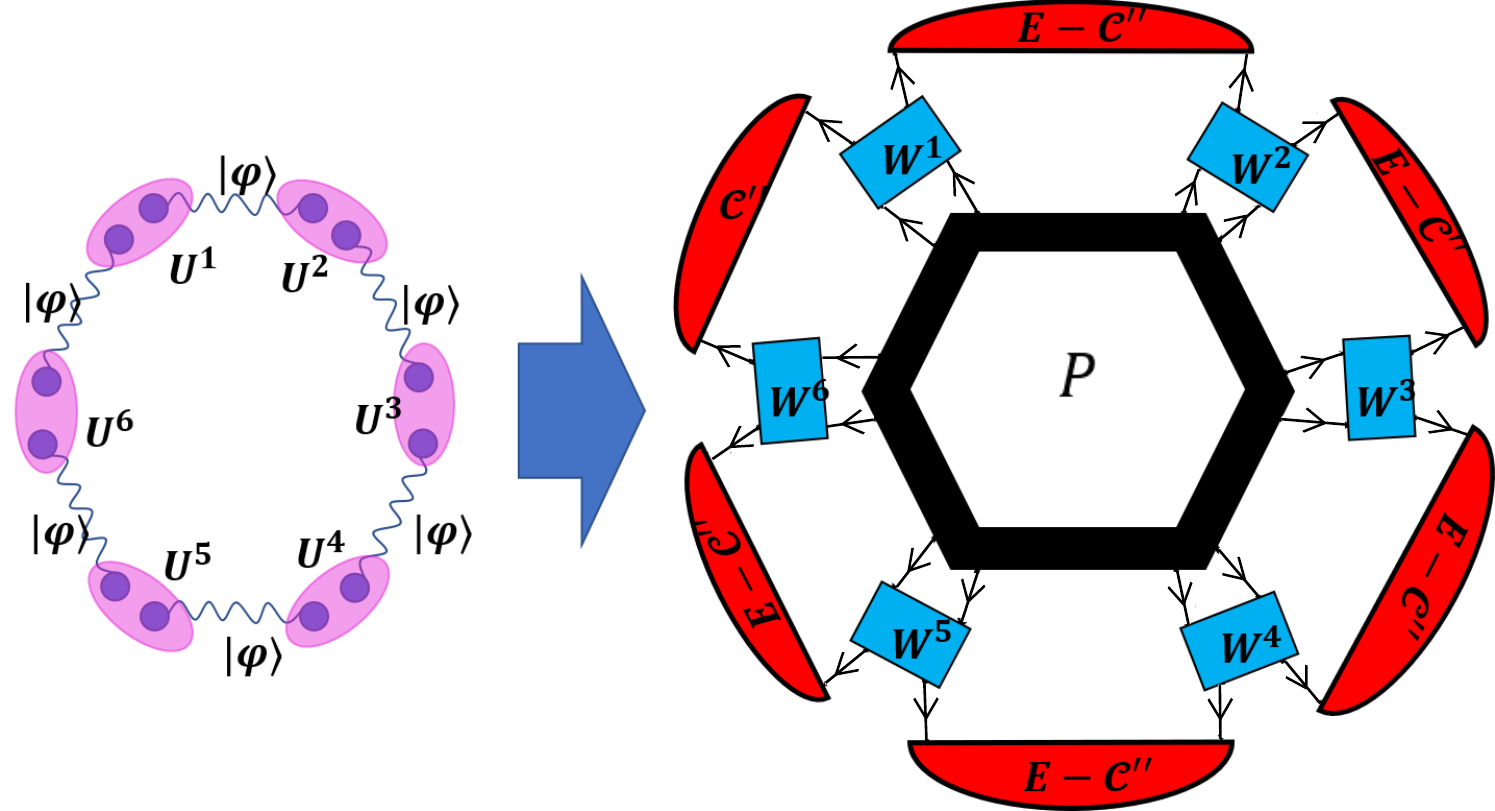}
  \caption{Left: a $2m$-partite quantum state built by acting with random two-qubit unitaries $U_1,...,U_m$ over many copies of the bipartite state $\ket{\phi}$. Right: contraction of $2\to 2$ and $2\to 1$ connectors used to detect the Bell nonlocality of the associated $2m$-partite box (in black).}
  \label{exNLPic}
\end{figure}

We find that connector theory provides a simple heuristic that relates the detection of nonlocality in a particular quantum experimental setup with the actual preparation of the underlying quantum state. Consider a scenario where several copies of the maximally entangled state $\ket{\phi}\equiv (e^{\frac{i\pi\sigma_y}{8}}\otimes \id_2)\frac{1}{\sqrt{2}}(\ket{0,0}+\ket{1,1})$ are acted with random two-qubit unitaries $U^1,...,U^m$ see \fref{exNLPic} (left). The resulting state---sometimes called a \textit{finitely correlated state} \cite{Werner}---is distributed among $2n$ parties, who probe it with Pauli measurements, thus obtaining a $2m$-partite non-signalling box $P$ with three inputs and two outputs at each site. Our goal is to certify that the $P$ is Bell nonlocal. 

Denote by $P'$ the three-input/two output box that results when we distribute $\ket{\phi}$ among two parties and allow each to measure its qubit with Pauli operators. Since Pauli measurements form a complete operator basis, we can identify any two-qubit operator $U$ with the way it transforms linear combinations of $\sigma_i\otimes \sigma_j$. It follows that there exist matrices $V^1,...,V^m$ such that $P=(V^1_{2,3}\otimes V^2_{4,5}...\otimes V^m_{2m,1}) (P')^{\otimes m}$. On the other hand, $P'$ violates one of the forms ${\cal C}''$ of the normalized CHSH Bell inequality when each party measures with $\sigma_x,\sigma_z$; more specifically, ${\cal C}''(P')=\frac{1}{2}-\frac{1}{\sqrt{2}}<0$. It follows that 

\begin{align}
&\{{\cal C}''\otimes (E-{\cal C}'')\otimes...\otimes (E-{\cal C}'')\} \nonumber\\
&([V^1_{2,3}]^{-1}\otimes ...\otimes [V^m_{2m,1}]^{-1})P=\nonumber\\
&\left(\frac{1}{2}-\frac{1}{\sqrt{2}}\right)\left(\frac{1}{2}+\frac{1}{\sqrt{2}}\right)^{m-1}.
\label{heurist}
\end{align}

Unfortunately, that does not prove that $P$ is Bell non-local, since $(V^1)^{-1},(V^2)^{-1},...$ are not connectors (that is, each one of them does not necessarily map classical boxes to classical boxes). Suppose, though, that we identified $2\to 2$ connectors $W^1, W^2,...$ whose action on $P$ were analogous to that of $(V^1)^{-1},(V^2)^{-1},...$, see \fref{exNLPic} (right). Then, there would be a fair chance that the newly devised Bell inequality

\begin{align}
B\equiv&\{{\cal C}''\otimes (E-{\cal C}'')\otimes...\otimes (E-{\cal C}'')\}\circ\nonumber\\
&(W^1_{2,3}\otimes ...\otimes W^n_{2n,1})
\end{align}

\noindent were such that $B(P)<0$.

To find a guess for $W^1$, we consider the box $Q=(\id_1\otimes V^1_{23}\otimes \id_4)P'\otimes P'$. It is easy to see that identifying the connector $W^1$ that minimizes $\{{\cal C}''\otimes (E-{\cal C}'')\} (\id_{1}\otimes W_{23}^1\otimes\id_4)Q$ can be cast as a linear program. Heuristically, $W^1$ is approximately inverting the action of $U^1$. Next, we consider the problem of identifying the connector $W^2$ such that $\{{\cal C}''\otimes (E-{\cal C}'')\otimes (E-{\cal C}'')\} (\id_1\otimes W^1V^1\otimes W^2V^2\otimes \id_6)(P'^{\otimes 3})$ is minimized; again an LP. We iterate this procedure until we obtain suitable guesses for $W^1,...,W^{m-1}$. The last step is to identify the connector $W^n$ that minimizes the contraction shown on the right side of \fref{exNLPic}. If the result is negative, we have detected the Bell nonlocality of $P$.

To assess how well this method works, we generated $300$ $m$-tuples of random unitaries $(U^1,...,U^m)$ for different values of $m$ and applied the procedure to the resulting $2m$-partite box. The results are shown in Table \ref{resChain}, Method I. Note that, for $m\geq 5$, the algorithm detected nonlocality \emph{always}.

\begin{table}
\begin{tabular}{c|c|c}
$m$& Method I &Method II\\
\hline
2 & 120 & 120\\ 
3 & 201 & 52\\ 
4 & 282 & 59\\
5 & 300 & 56\\
6 & 300 & 65\\
\end{tabular}
\caption{Cases (out of $300$) where the nonlocality of the chain state in \fref{exNLPic} (left) was detected via the connector contraction in \fref{exNLPic} (right) for two different methods to choose the connectors. The numbers in the first row are the same because both methods are equivalent for $m=2$.}
\label{resChain}
\end{table}

Notice that, if the intuition behind the heuristic is taken literally, the detection of Bell nonlocality should only depend on the values of just $U^1,U^2,U^n$. Indeed, intuitively, the ${\cal C}''$ red connector in \fref{exNLPic} is associated to the first negative term on the right-hand side of eq. (\ref{heurist}): the remaining red connectors $E-{\cal C''}$ give the positive contribution on the right of the equation; and are just meant to enhance the magnitude of the Bell violation. This intuition leads one to anticipate that, for $m\geq 3$, the probability of a Bell violation should not depend on the system size. In fact, we observe just the opposite: as the system grows in size, the probability of detecting non-locality with the contraction in \fref{exNLPic} increases with $m$, see Table \ref{resChain}.

A possible explanation is that, as the index $k$ runs from $1$ to $n$, the action of $W^k$ becomes less and less the inversion of $V^k$. On the contrary, the heuristic seems to be exploiting the structure of the correlations between the remaining parties after the application of $\{{\cal C}''\otimes (E-{\cal C}'')^{\otimes k}\}$ in order to boost the overall Bell violation even further. Actually, if we modify the heuristic and, for $k\geq 2$, we derive each $W^k$ by maximizing the contraction

\begin{align}
&\left(E\otimes (E-{\cal C})\otimes (E-{\cal C})\otimes E\right)\nonumber\\
&\left(\id_1\otimes W^{k-1}\otimes W^{k}\otimes \id_{6}\right)(P')^{\otimes 3},
\end{align}
\noindent then the dependence in $m$ disappears, see Table \ref{resChain}, Method II.

\subsubsection{Nonlocality detection in more general boxes} 
\begin{figure}
  \centering
  \includegraphics[width=6.5cm]{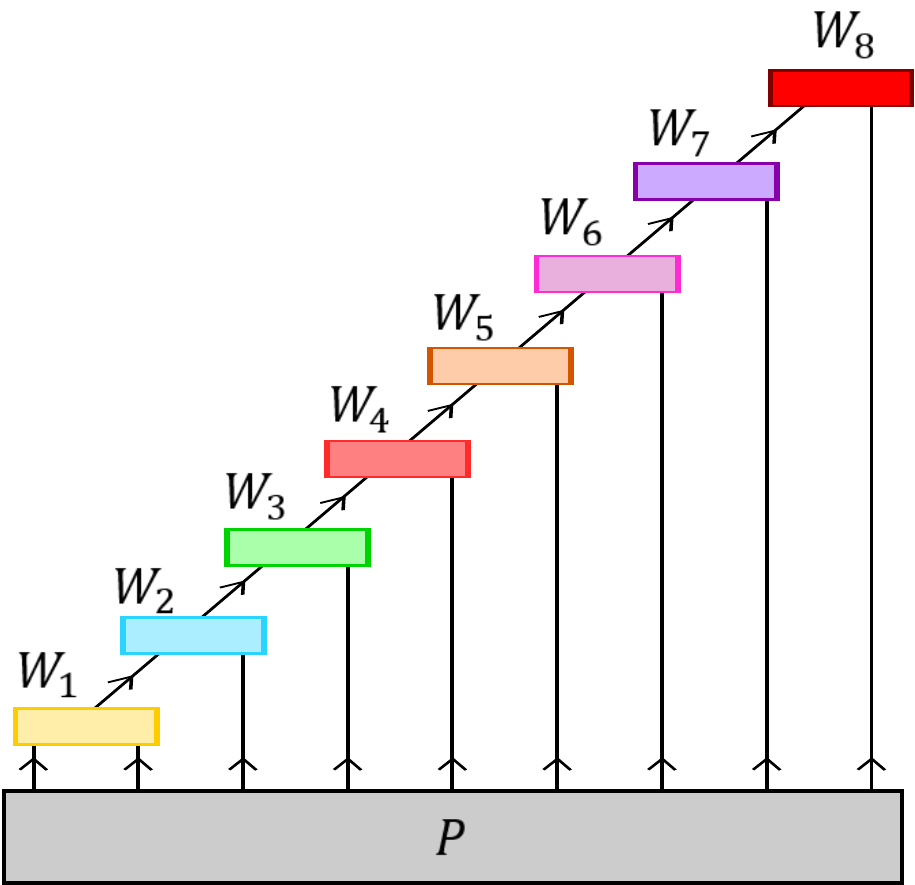}
  \caption{\textbf{A Matrix Product Connector Tensor Network (MPCTN).} A generic witness composed from $2 \to 1$ connectors that are contracted similar to tensors in a matrix product states (see \fref{fig:TN}(b)). We found the resulting witness---the MPCTN---useful for detecting non-locality and entanglement in several systems. The sequence of $2 \to 1$ connectors applied from left to right coarse-grain the box to an effective 2-site box, whose global properties can be explored by means of a 2-site witness $W_8$. When $P$ also has an efficient tensor network representation, then an MPCTN provides a witness that can be scaled to hundreds of sites.}
  \label{MPSPic}
\end{figure}

In some situations, we want to decide the Bell nonlocality of a multipartite box for which one just has a theoretical description. (That is, no preparation information is available as in the previous discussion.) To address nonlocality detection in more general boxes, we introduce the \textit{Matrix Product Connector Tensor Network} (MPCTN): a witness composed of $2 \to 1$ connectors that are contracted similar to tensors in a matrix product state, see \fref{fig:TN}(b). \fref{MPSPic} illustrates a MPCTN for a 9-party box.

Denote by $n_I$ and $n_O$ the number of inputs and the number of outputs that appear on the bond indices of the MPCTN respectively. We call the pair $(n_I,n_O)$ the \textit{bond dimension} of the MPCTN. The bond dimension determines the size of the output index of each connector. In principle, one can choose a different bond dimension for each connector. However, in the numerical simulations presented here, we fixed the same bond dimension for all the connectors in the MPCTN.
We remark that even though in this paper all numerical results were obtained by using a MPCTN witness, we can at least construct a MERA-like witness, since non-trivial $2\to 2$ connectors exist (e.g. the one depicted in \fref{disentPic}).

Having thus fixed MPCTN as the ansatz for non-locality witness, the next task to (numerically) determine the connectors connectors $\{W_i\}$ in order to minimize the contraction shown in the figure for a given box $P$. In numerical optimizations, the bond dimension controls both the computational cost and the value of possible violations. We used two types of optimization techniques.

\textbf{See-saw optimization.---} Let us initialize all the connectors to random values (within the space of $2\to 1$ connectors of the given bond dimension). Denote by $W(P)$ the value of the contraction illustrated in \fref{MPSPic}. We can fix all but one connector, say at location $j$, and determine connector $W^j$ that minimizes $W(P)$. This problem reduces to a linear optimization over the set of $2\to 1$ connectors, a problem that can be cast as linear program.

Iterating, we obtain a sequence of decreasing values for $W(P)$. We can stop the protocol as soon as $W(P)$ becomes negative. This sort of optimization procedures are called \emph{see-saw methods} \cite{seeSaw1,seeSaw2}, and they have proven very helpful in condensed matter physics and quantum nonlocality. In our numerical simulations, though, we find that, unless $W(P)$ is negative from the very beginning, very often one of the optimal connectors becomes $0$. In such cases, a projected gradient method \cite{subgradient} seems to be a better choice to minimize $W(P)$. 

\textbf{Gradient descent optimization.---}Choose $\epsilon>0, \epsilon\ll 1$, and let $E_j$ denote the tensor obtained by contracting all tensors except $W_j$. We will say that $E_j$  is the `environment' of tensor $W_j$. Adapted to this problem, the subgradient method consists in updating the connectors via the iterative equation:

\be
\Omega^{k+1}_j=\pi_C\left(\Omega^k_j-\epsilon E^k_j\right).
\ee
\noindent Here, for any tensor $A$, $\pi_C(A)$ denotes the projection onto the set of valid connectors. That is, $\pi_C(A)$ is the connector that best approximates $A$ in $2$-norm (when viewed as a multipartite vector). Computing projections can be formulated as an SDP, and hence it can be solved efficiently, as long as the cardinality of the indices of the connector is kept at a reasonable value.

\begin{figure}
  \centering
  \includegraphics[width=7.5cm]{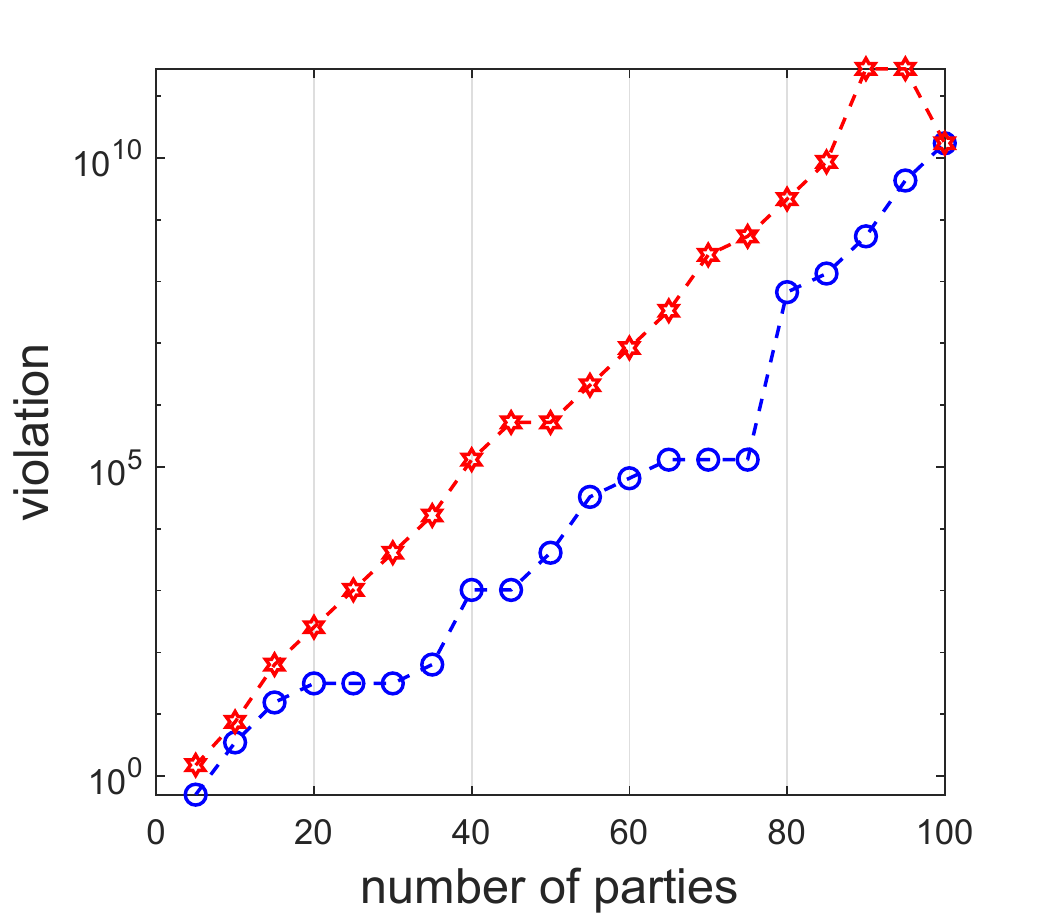}
  \caption{\textbf{Bell violations for the GHZ state with fixed measurements for number of parties = $\{5,10,15,...,100\}$.} Bond dimension = (2 2). The blue points (circles) are the violations obtained using the see-saw optimization scheme using randomly chosen initial connectors until 75 parties. Using random initial connectors failed to produce violations for larger number of parties. The violations for larger sizes (between 75 and 100) were obtained by using the \textit{optimized} connectors for $L$ parties as the initial guess for the optimization of $L+5$ parties. The red points (star) are substantially enhanced violations obtained by using the \textit{optimized} connectors for 100 parties as the initial guess for smaller number of parties, suggesting that a simple see-saw optimization scheme may not be optimal. (The 100 party solution stopped working as the initial guess for less than 35 parties. We still managed to obtained enhanced violations in this domain by feeding the optimized connectors for $L$ parties as the initial guess for $L-5$ parties.)}
  \label{GHZViolations}
\end{figure}

\textbf{Guess for initial connectors.---} While some times using random connectors as the initial guess for the optimization (both the see-saw and the gradient descent) worked well, we observed that in some cases making an educated guess for the initial connectors produced a violation when starting with random initial connectors failed to do so. (This also some times enhanced the violation in cases where there was a violation with random initial connectors.) A guess that often worked when exploring the same box for larger and larger number of parties was to use the \textit{optimized} connectors for smaller box as initial connectors for the larger number of parties. Some times, the reverse worked---optimized connectors for larger number of parties provided a good guess for smaller number of parties. In practice, we tried such different schemes to determine a method that worked well for a given box.

\begin{figure}
  \centering
  \includegraphics[width=8cm]{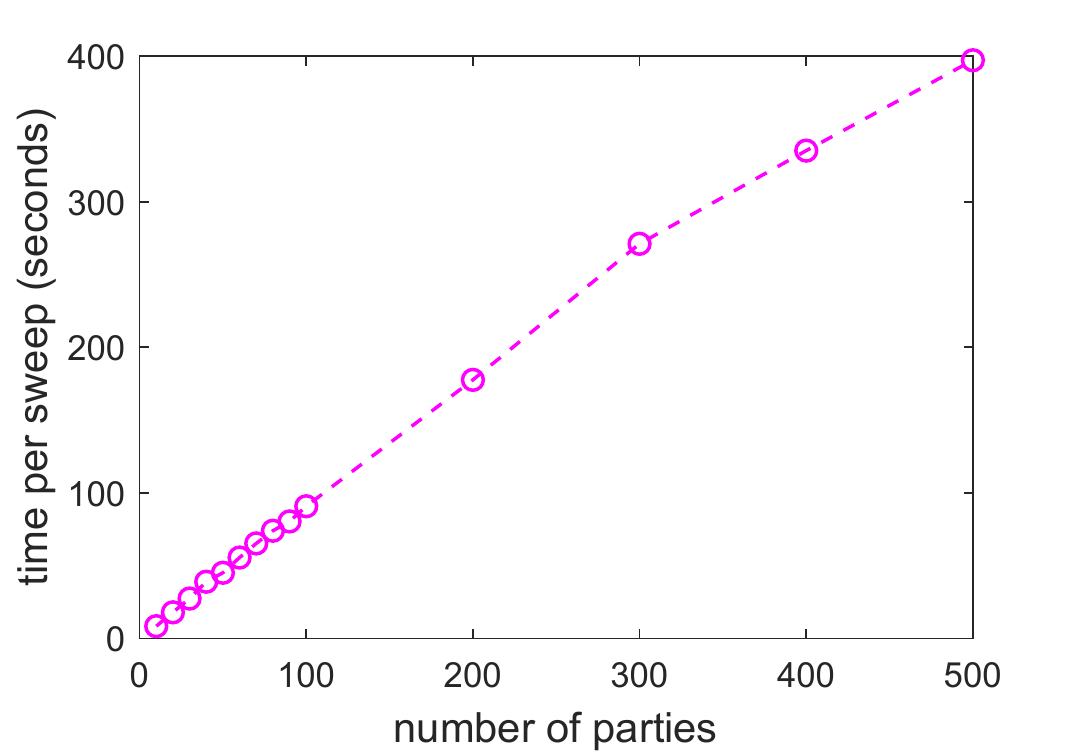}
  \caption{\textbf{Scalability.} Computational time (in seconds) for a single see-saw sweep for optimization of bell nonlocality connectors with bond dimension = (2,2) grows approximately linearly with number of parties. Simulations were run on a regular laptop with 2.5 GHz CPU and 8 GB RAM.} 
  \label{GHZViolations2}
\end{figure}
\textbf{Scalability.---} There exist relevant scenarios in which the no-signalling box $P$ can also be efficiently represented as a tensor network e.g. a matrix product state of a low bond dimension. This is possible when the box is the result of measuring a quantum state with limited correlations, for example, the thermal state of a 1D local Hamiltonian. In this case, the contraction illustrated in \fref{MPSPic} to compute $W(P)$ can be carried out with a cost that scales only \textit{linearly} with the number of parties \cite{cirac}. This means that we can apply the method above to assess the Bell nonlocality of boxes shared by hundreds or even thousands of parties. However, note that increasing the bond dimension---either of the box $P$ or of the MPCTN--- increases the pre-factor in the scaling of the computational cost.

\textbf{Benchmarking result: GHZ state.---}To test how these method works, we considered the quantum box that results when $m$ parties share the GHZ state \cite{GHZ} $\ket{\psi}=\frac{1}{\sqrt{2}}(\ket{0}^{\otimes m}+ \ket{1}^{\otimes m})$ and are allowed to measure it with the settings $\{\frac{\id\pm\sigma_x}{2},\frac{\id\pm\sigma_z}{2}\}$. The result is a box which can be expressed as an MPS of bond dimension 4. 

Figure \ref{GHZViolations} shows the violations we obtained for $m = \{5,10,15,...,100\}$ using two see-saw optimization schemes. Note that the magnitude of the violation seems to increase approximately exponentially with the number of parties. (In fact, we were unable to proceed for number of parties much larger than 100 because of instabilities in optimizing the connectors owing to the presence of very large coefficients in connector tensors.) This rules out the possibility that the optimization algorithm is simply determining wirings to project the last three parties into a GHZ state and then probing the latter with, e.g., the Mermin inequality \cite{mermin}. The algorithm is finding a cleverer solution. Furthermore, we found violations only after a couple of see-saw sweeps, even when starting from random connectors. Note that detecting the non-locality of a system made of hundreds parties, as done here, is completely out of reach with the existing techniques.

\section{Supraquantum nonlocality detection}
\label{supraQ}
In this section, we sketch how to use connector theory to determine whether a given non-signalling box $P_{a_1,...,a_m}^{x_1,...,x_m}$ admits a quantum realization. The object of interest is the same as above, a probability distribution for the measurement outputs conditioned on the inputs, but now the goal is to understand whether these given correlations can be reproduced within the quantum formalism. That is, whether there exist Hilbert spaces $\H_1,...,\H_m$ and operators $\{M^k_{a|x}:\H_k\to\H_k:k=1,...,m,a=1,...,d,x=1,...,n\}$, $\rho:\H_1\otimes...\H_m\to \H_1\otimes...\H_m$ such that

\begin{align}
P_{a_1,...,a_m}^{x_1,...,x_m}=\tr\left\{\rho_{1,...,n} (M^1)_{a_1|x_1}\otimes...\otimes M^m_{a_m|x_m})\right\},
\end{align}

\noindent where $\rho$ is a non-normalized quantum state and $M^k_{a|x}$ is to be understood as the Positive Operator Valued Measure (POVM) element corresponding to party $k$ inputting $x$ in its box and obtaining the result $a$. That is,
\be
M^k_{a|x}\geq 0, \sum_{a}M^k_{a|x}=\id_{\H_k}.
\label{POVM}
\ee

The natural GPT to consider here would be \textbf{QUANT-world}, whose states are quantum boxes of arbitrarily many inputs and outputs. In this case, the dual set ${\cal Q}'$ of the set of quantum boxes corresponds to the set of coefficients $W^{a_1,..,a_n}_{x_1,..,x_n}$ such that, given any set of operators $\{M^k_{a|x}\}$ satisfying (\ref{POVM}), the operator

\be
\sum_{a_1,...,a_n,x_1,...,x_n}W^{a_1,..,a_n}_{x_1,..,x_n}M^1_{a_1|x_1}\otimes...\otimes M^n_{a_n|x_n}
\label{positivo}
\ee
\noindent is positive semidefinite. As explained in Appendix \ref{appQuant}, non-trivial SDP ans\"atze on ${\cal Q}'$ can be derived from the Navascu\'es-Pironio-Ac\'in hierarchy (NPA) \cite{npa,npa2,dohertyLiang} and variants \cite{berta}.

We claim that, replacing ${\cal B}'$ in Eq. (\ref{coarseB}) by ${\cal Q}'$, we arrive at a characterization of the set of $m\to 1$ transformations between quantum boxes. Indeed, consider a general transformation $W$ such that $\{W_{c|z}:c,z\}$, acting on the POVM elements $\{M^k_{a|x}\}$ via (\ref{positivo}), generate positive semidefinite operators $N_{c|z}$ with the property that $\sum_{c}N_{c|z}=N\leq \id$. When applied over the box $P_{a_1,...,a_{m+r}}^{x_1,...,x_{m+r}}$, the resulting $r+1$-partite box $P_{c,a_{m+1},...,a_{m+r}}^{z,x_{m+1},...,x_{m+r}}$ admits the decomposition

\begin{align}
&P_{c,a_{m+1},...,a_{m+r}}^{z,x_{m+1},...,x_{m+r}}=\nonumber\\
&\tr\left\{\tilde{\rho} (\tilde{N}_{c|z}\otimes M^{m+1}_{a_{m+1}|x_{m+1}}\otimes...\otimes M^{m+r}_{a_{m+r}|x_{m+r}})\right\},
\label{reform}
\end{align}
\noindent where $\tilde{N}_{a|x}\equiv N^{-1/2}N_{a|x}N^{-1/2}$ and $\tilde{\rho}=(N^{1/2}\otimes \id_{m+1,...,m+r})\rho_{1,...,m}(N^{1/2}\otimes \id_{m+1,...,m+r})$. Note that $\{\tilde{N}_{c|z}\}_c$ can be interpreted as the POVM elements of an $m$-partite global measurement $z$ with outcome $c$ on the post-selected state $\tilde{\rho}$. The latter, in turn, is the result of effecting the trace-decreasing map $N^{1/2}$ on part of $\rho$. Hence $P_{c,a_{m+1},...,a_{m+r}}^{z,x_{m+1},...,x_{m+r}}$ defines an $r+1$-partite quantum box.

To test the method, next we introduce a number of non-signalling supra-quantum boxes which admit an MPS decomposition of bond dimension linear on the system size, or even bounded. This will allow us to test their non-quantumness for high system sizes.

\subsection{Generalized Svetlichny box}
Consider a scenario where $m$ parties have two measurements, each with two outcomes, i.e., $x_1,...,x_m,a_1,...,a_m\in\{0,1\}$, and let $f(x_1,...,x_m)$ be any Boolean function. It can be verified that the box with statistics 

\be
P_{a_1,...,a_m}^{x_1,...,x_m}=\frac{1}{2^{n-1}}\delta_{\bigoplus_k a_k,f(x_1,...,x_m)}
\label{funda}
\ee
\noindent is normalized and no-signalling. The proof is simple: if we trace out any party, the probability of any sequence of outputs equals $1/2^{m-1}$, independently of the sequence of inputs. Almost all such boxes allow, by wirings, to produce a perfect PR box, and hence they cannot be realized within quantum theory. Moreover, there exist important supra-quantum boxes within this family, such as the Svetlichny box \cite{svetlichny}. Next we generalize the Svetlichny box to arbitrarily many parties and then show that it admits an MPS representation with bond dimension $16$.

The original Svetlichny box is tripartite, with statistics given by
\be
P_{a_1,a_2,a_3}^{x_1,x_2,x_3}=\frac{1}{2^2}\delta_{a_1\oplus a_2\oplus a_3,x_1x_2\oplus x_2x_3\oplus x_3x_1}.
\ee
It is thus of the form (\ref{funda}), with $f(x_1,x_2,x_3)=x_1x_2\oplus x_2x_3\oplus x_3x_1$. We will generalize it to a box of the form (\ref{funda}), with $f(x_1,...,x_m)=\bigoplus_{k=1}^{m-1}x_kx_{k+1}\oplus x_m x_1$. It can be verified that all such `generalized Svetlichny boxes' can be simulated by distributing a PR-box to each party and its near neighbor and let each party wire its two boxes together. Again, reaching similar results for systems of hundred particles is impossible with existing methods.

\begin{figure}
  \centering
  \includegraphics[width=7cm]{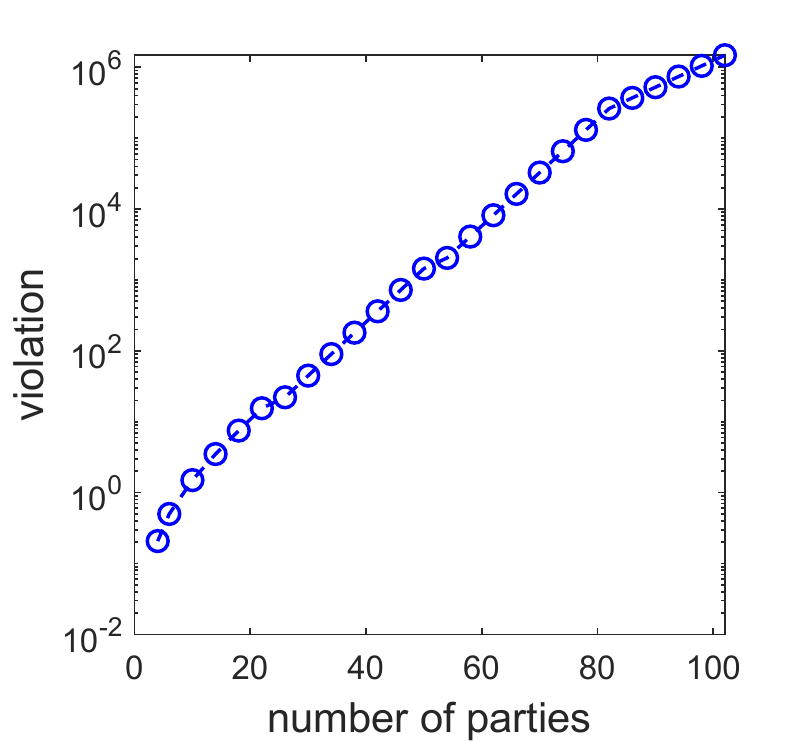}
  \caption{\textbf{Quantum nonlocality violation of the Svetlichny box for number of parties = $4,8,12,...,102$}. Maximum bond dimension = (4 4). The violations were obtained using the see-saw optimization scheme. We used the \textit{optimized} connectors for $L-4$ parties as the initial guess for connectors for $L$ parties.}
  \label{svetPic}
\end{figure}

In turn, Svetlichny boxes can be seen to admit an MPS decomposition 

\be
P_{a_1,...,a_m}^{x_1,...,x_m}=\Lambda^{[1]}_{a_1,x_1}...\Lambda^{[m-1]}_{a_{m-1},x_{m-1}}\Lambda^{[m]}_{a_m,x_m}
\label{MPS}
\ee
\noindent involving matrices $\Lambda^{[k]}_{a,x}$ of size at most $16\times 16$, see Appendix \ref{appNSBoxes} for their exact expression.

Using the above MPS representation of the Svetlichny boxes and by using a MPCTN witness composed of $2 \to 1$ connectors from \textbf{QUANT-world}, we were able to detect quantum nonlocality violation in these boxes for large number of parties, see \fref{svetPic}. We find that violation increases exponentially with the number of parties.

\subsection{Other boxes}
Another family of boxes is obtained by defining 
\begin{align}
f_r(x_1,...,x_m)=&1, \mbox{ if $x_1,...,x_m$ contains $r$ consecutive $1s$},\nonumber\\
&0, \mbox{ otherwise}.
\label{consecBox}
\end{align}
\noindent This box also admits a MPS representation (\ref{MPS}) for bond dimension $2(r+1)$, see Appendix \ref{appNSBoxes}.

A fully symmetric `majority voting' box is given by the function
\begin{align}
\mbox{maj}(x_1,...,x_m)=&1, \mbox{ if half or more of the inputs are $1s$},\nonumber\\
&0, \mbox{ otherwise}.
\label{majBox}
\end{align}
\noindent This time, the bond dimension of the box scales linearly with the system size $m$.

\begin{figure}
  \centering
  \includegraphics[width=7cm]{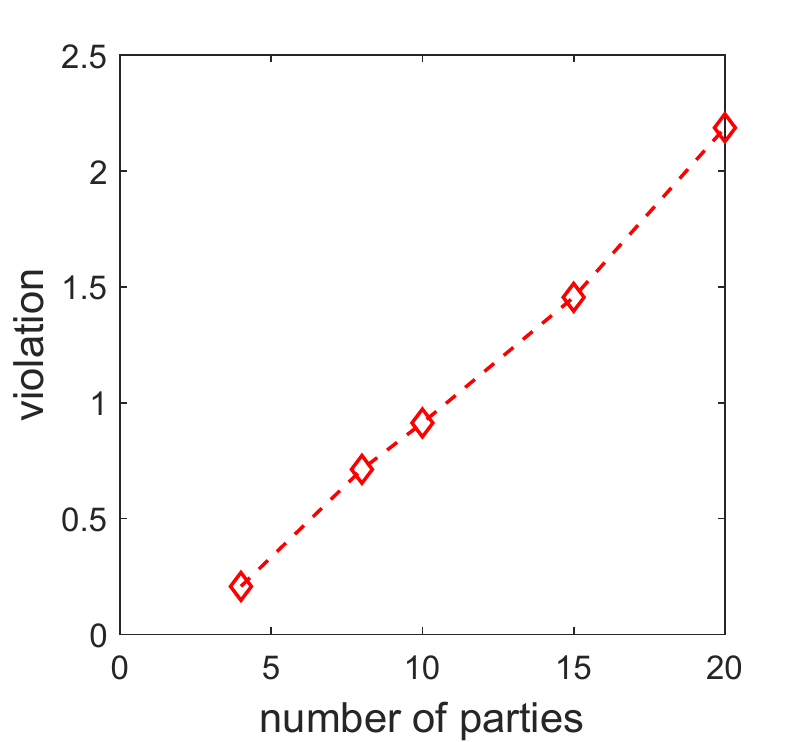}
  \caption{\textbf{Quantum nonlocality violation of the box defined in \eref{consecBox} with $r=2$.} Maximum bond dimension = (4 4). With the optimizations techniques described in this paper we managed to find violations only up to 20 parties. It is possible that violations will also be found for more parties by using more sophisticated see-saw optimization techniques, better guess for initial connectors, and/or using larger bond dimensions. With our current implementation, we could only manage to run simulations with a maximum bond dimension = (4 4) in a reasonable time.}
  \label{m2Pic}
\end{figure}

Figure \ref{m2Pic} shows the violations we found for the box defined in \eref{consecBox} with $m=2$, for up to 20 parties. We also tried the same box with $m=3$, but managed to find a small violation $\approx -0.1$ for 4 parties. Even for this case, we used a bond dimension = (4,4). The optimization of the connectors with larger bond dimension is very slow.

\section{Entanglement detection}
\label{secEnt}
As a final application of connector theory, we come back to the problem of entanglement detection, which we used in the introduction to illustrate the main intuition of our construction. 
To apply connector theory to detect entanglement, we define a GPT whose states coincide with the fully separable quantum states, call it \textbf{SEP-world}. In this theory, the norm of a state $\rho$ is defined as $E(\rho)\equiv \tr(\rho)$. As in \textbf{LOC-world}, due to the structure (\ref{sepState}) of the set of separable states, \textbf{SEP-world} has the property that any linear map fulfilling condition (\ref{NRH}) with $T=\emptyset$ constitutes a valid transformation. The set of connectors in \textbf{SEP-world} thus corresponds to the set of linear maps which transform separable states into separable states. In general, these operations cannot be implemented in quantum theory.

When studying \textbf{LOC-world}, we noted that the structure of $m\to 1$ connectors is closely linked to that of Bell inequalities. Similarly, in \textbf{SEP-world} there is a one-to-one correspondence between scaled connectors and $m+1$-partite entanglement witnesses \cite{EW}. We remind the reader that a $k$-partite entanglement witness $W$ is an operator acting in $\bigotimes_{i=1}^k\H_i$ such that

\be
\tr(W\rho)\geq 0,
\ee
\noindent for all fully separable states $\rho$. 

We claim that a linear map $\Omega:B(\otimes_{i=1}^m\H_i)\to B(\H_{m+1})$ corresponds to an $m\to 1$ connector iff:

\begin{enumerate}
\item
$W_\Omega$, defined via the relation $\tr\{W_\Omega(\sigma\otimes \beta)\}=\tr\{\Omega(\sigma)\beta\}$, is an $m+1$-partite entanglement witness,
\item
$\id_{1,...,m}-\tr_{m+1}(W_\Omega)$ is an $m$-partite entanglement witness.
\end{enumerate}

Let us see why this correspondence holds. Let $W$ be an $m+1$-partite entanglement witness; and $\rho$, any separable state of the form $\rho=\sigma_{1,...,m}\otimes\beta$. Consider the map $\Omega_W(\sigma)=\tr_{1,...,m}\{(\sigma\otimes \id_{m+1})W\}$. By definition, we have that $\tr(\Omega(\sigma)\beta)=\tr(W\rho)\geq 0$. Since $\beta\geq 0$ is arbitrary, this means that $\Omega_W(\sigma)\geq 0$, for all fully separable states $\sigma$. That is, $\Omega_W$ is a scaled connector. Conversely, any linear map $\Omega$ with the property that $\Omega(\sigma)\geq 0$ for all fully separable states $\sigma$ can be mapped to an entanglement witness $W_\Omega$ as defined in condition 1. This establishes that $\Omega$ is a scaled connector iff condition 1 holds. Condition 2 is easily seen to ensure that the norm is non-increasing, i.e., $\tr\{\Omega(\sigma)\}\leq \tr\{\sigma\}$.

The above observation allows us to link the connector theory of \textbf{SEP-world} with the existing literature in entanglement detection. In principle, we can promote any $k$-partite witness to a $k-1\to 1$ connector and contract several copies thereof, as we did with the CHSH Bell inequality in \fref{treePic}. The result would be a novel entanglement witness for $m$-partite entangled states. 

\begin{figure}
  \centering
  \includegraphics[width=3cm]{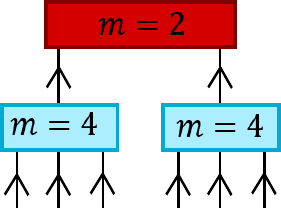}
  \caption{\textbf{Connectors for entanglement detection.} Six-qubit entanglement witness $W_6$ resulting from the composition of witness (\ref{geza}) with $m=2$ and $m=4$.}
  \label{EWPic}
\end{figure}

Take, for instance, the family of $m$-qubit entanglement witnesses derived by Toth et al. in \cite{spin}:

\be
\sum_{i=x,y,z}\left\langle(J_i-\langle J_i\rangle)^2\right\rangle\geq \frac{m}{2},
\ee
\noindent with $J_i=\frac{1}{2}\sum_{j=1}^m\sigma^{(j)}_i$. This is not a linear witness, but can be turned into one by just replacing $\langle J_i\rangle$ by arbitrary real numbers:

\be
\sum_{i=x,y,z}\left\langle(J_i-\lambda_i)^2\right\rangle\geq \frac{m}{2}.
\label{geza}
\ee
Taking $\lambda_i=0$, one can contract the connectors associated to the $4$ and $2$-qubit entanglement witnesses as shown in \fref{EWPic} to produce the witness $W_6$. Numerically, we find that there exist $6$-qubit states which, while satisfying all forms of (\ref{geza}), can be detected by $W_6$. This shows that new detection properties can arise from composition alone. We come back to this in Section \ref{hybrid}.

Constructing witnesses which detect the entanglement of a given quantum state is a more complicated task, due to the difficulty of certifying that $W_\Omega, \id_{1,...,m}-\tr_{m+1}(W_\Omega)$ are indeed entanglement witnesses. A possible approach to this problem is to prove instead that the average values of those two operators are non-negative when evaluated over a \emph{relaxation} (a superset) of the set of separable states. The family of relaxations which we considered in our numerical examples is called the Doherty-Parrilo-Spedaliery (DPS) hierarchy \cite{DPS1,DPS2,DPS3}. Combining this idea with the observation in \cite{NOP} that a small perturbation of the DPS sets projects them to the interior of the set of separable states, in Appendix \ref{appSep} we present a family of SDP ans\"atze on the set of $m\to m'$ connectors. Throughout the rest of this section, we use those ans\"atze whenever a linear optimization over feasible connectors is required.

\subsection{PPT states}
To test how useful connectors are for entanglement detection, we first considered a famous class of multipartite entangled states which are positive under partial trasposition (PPT) \cite{peres}. An unextendible product basis (UPB) is a collection of $m$-partite orthogonal product states $\{\ket{\psi_i}\}_{i=1}^K$ with the property that no other product vector is orthogonal to their span. Given any UPB, $\{\ket{\psi_i}\}_{i=1}^K$, the $m$-partite state

\be
\rho\propto \id_{1,...,m}-\sum_i\proj{\psi_i}
\label{PPTEnt}
\ee
\noindent can be shown entangled and PPT \cite{UPB1}. In \cite{UPB2}, a family of six-qubit UPBs parametrized by three qubit unitaries, is presented. We sampled $10$ unitary triples randomly according to the Haar measure, built the corresponding six-qubit quantum states (\ref{PPTEnt}), and used a MPCTN to detect their entanglement. The output system of each connector was a qubit. In all cases, a see-saw algorithm found a normalized entanglement witness whose average value of the state was $\simeq -0.5$.

\begin{figure}
  \centering
  \includegraphics[width=5cm]{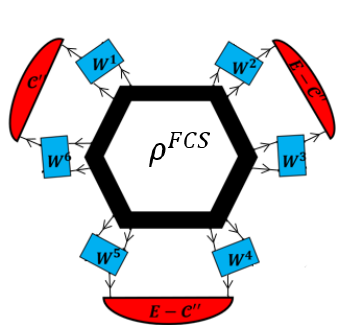}
  \caption{Contraction of $2\to 1$ connectors used to detect entanglement of a finitely correlated mixed state $\rho^{FCS}$. Here, $\mathcal{C}' = $ SWAP is the witness for certifying entanglement in the singlet state $\ket{\Psi^-}$.}
  \label{fig:PrepDetectionMixed}
\end{figure}

\subsection{Finitely correlated mixed states}

We generated mixed states following a preparation similar to the one described in Sec.~\ref{ssec:FCS}. We considered distributing $m$ singlets $\ket{\Psi^-} \equiv \frac{1}{2} (\ket{0}\ket{1} - \ket{1}\ket{0})$ amongst $2m$ parties as illustrated in \fref{exNLPic} (where state $\ket{\phi}$ is replaced with $\ket{\Psi^-}$). We also replaced the action of unitaries $U_1,U_2,...,U_m$, shown in the figure, by conjugation with a convex combination of two unitaries (drawn randomly for each pair of sites). The resulting state is mixed, has an efficient MPS representation, and may be separable. We wanted to certify whether such states are entangled or not using connectors.

We used a witness to the one illustrated in \fref{exNLPic} (right hand side). The intuition is the same. We wanted to find $2\to2$ connectors in \textbf{SEP-world} that approximately inverted the randomizing quantum channel, thus exposing the initially singlets. The singlets can be certified to be entangled by a 2-party witness which is simply the SWAP gate, which evaluates to $-1$ for the singlet $\ket{\Psi^-}$. (That is, we replaced $\mathcal{C}' = $SWAP in \fref{exNLPic}; $E-$SWAP connectors are once again used to amplify the violation.)

\begin{figure}
  \centering
  \includegraphics[width=6cm]{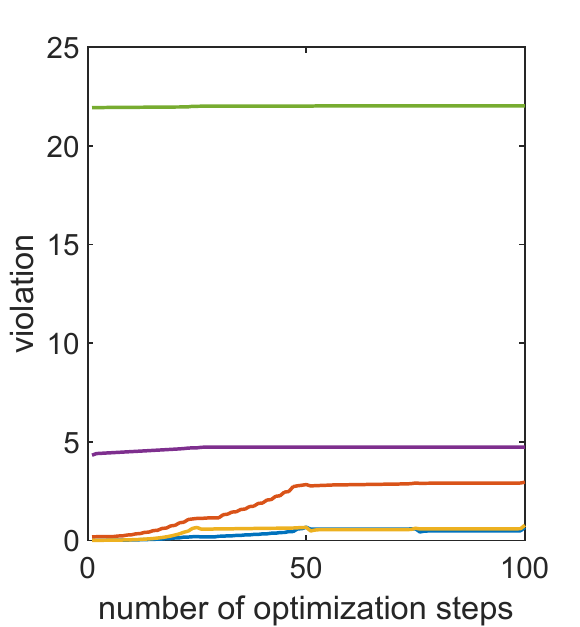}
  \caption{\textbf{Entanglement detection in FCS mixed states of 50 qubits using $2\to2$ connectors.} We randomly sampled five states and found a violation in each case. Each optimization step consists of optimizing one connector.}
  \label{fig:FCSEntanglement}
\end{figure}

We generated such finitely correlated mixed states from randomly chosen unitaries (using the Haar measure) for a system of 50 qubits, and found a violation (certifying the presence of entanglement) almost each time. \fref{fig:FCSEntanglement} shows the violations obtained for five such randomly drawn states.

\begin{figure}
  \centering
  \includegraphics[width=6cm]{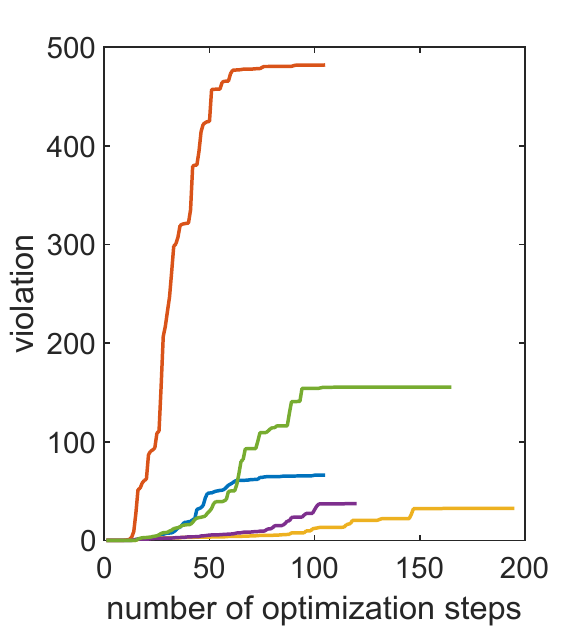}
  \caption{\textbf{Entanglement detection in FCS mixed states of 60 qubits using $2\to1$ connectors..} We randomly sampled five states and found a violation in each case. Each optimization step consists of optimizing one connector. We found much larger violations than when using $2\to2$ connectors.}
  \label{fig:FCSEntanglement1}
\end{figure}

We also used another witness, one composed of only $2\to1$ connectors as illustrated in \fref{fig:PrepDetectionMixed}. Again, we easily found violations for randomly chosen states, see \fref{fig:FCSEntanglement1}. We also found that the violations were larger than those obtained by using $2\to2$ connectors.

\subsection{Entanglement detection through hybrid GPTs}
\label{hybrid}

In the previous section, we linked the entanglement problem to connector theory by defining a GPT, \textbf{SEP-LOC}, where the set of physical states coincides with the set of fully separable states. The purpose of this section is to demonstrate that it is even possible to follow a hybrid approach in which different GPTs are connected. 

Consider, for example, a theory where there are two types of basic systems: quantum systems and boxes. The state of a composite system of, say, two boxes and two quantum systems, would be a steering ensemble of the form $\{\rho^{a|x,b|y}_{CD}\}$, with the property that there exists a fully separable quantum state $\sigma_{ABCD}$ and measurement operators $M^A_{a|x},M^B_{b|y}$ such that 

\be
\rho^{a|x,b|y}_{CD}=\tr_{AB}\{\sigma M^A_{a|x}\otimes M^B_{b|y}\otimes\id_{CD}\}.
\ee
\noindent We dub this theory \textbf{STEER-world}.

\begin{figure}
  \centering
  \includegraphics[width=2.5cm]{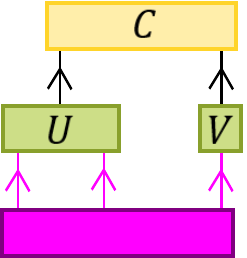}
  \caption{\textbf{Entanglement detection through hybrid GPTs.} Purple lines indicate quantum systems; black lines, boxes. Starting from a quantum state, we effect two transformations $U,V$ to map it to a bipartite box, which we then probe with the normalized CHSH inequality.}
  \label{mixedPic}
\end{figure}

Now, suppose that we wished to assess the entanglement of a three-qubit state. One possibility would be to regard it as a possible state of \textbf{STEER-world} and then apply the connectors depicted in \fref{mixedPic}. There the three-partite quantum state is transformed into a bipartite box, which we then evaluate with the normalized CHSH inequality ${\cal C}$ (\ref{CHSHs}). This scenario reminds that of device-independent certification of entanglement, and actually it would be equivalent, if the transformations $U,V$ acted on single systems. Indeed, in that case connectors from quantum systems to boxes correspond to conducting quantum measurements on the former, and entanglement is detected iff the corresponding box violates a Bell inequality.

As we will see, the $2\to 1$ connector $U$ mapping bipartite quantum systems to a single box changes things completely. Let $U,V$ be defined via:

\begin{align}
&U^{0|0}(\rho)=\tr(SWAP\rho), \nonumber\\
&U^{0|1}(\rho)=\tr\{\rho\frac{1}{2}(\id_4+\sigma_y\otimes\sigma_y)\},\nonumber\\
&V^{0|0}(\beta)=\tr\{\beta\frac{\id_2+\sigma_x}{2}\}, \nonumber\\
&V^{0|1}(\beta)=\tr\{\beta\frac{\id_2+\sigma_z}{2}\},
\end{align}
\noindent where the superindex in each tensor corresponds to the index of its upper, black leg. Here $SWAP$ denotes the permutation operator $\sum_{i,j=0,1}\ket{i}\bra{j}\otimes\ket{j}\bra{i}$; and $\{\sigma_i\}_{i=x,y,z}$, the three Pauli matrices. We assume that $U,V$ are deterministic transformations, i.e., $U^{1|x}(\rho)=E(\rho)-U^{0|x}(\rho),V^{1|y}(\beta)=E(\beta)-V^{0|y}(\beta)$, for $x,y=0,1$.

Note that $U^{0|1}$, $V^{0|0}$, $V^{0|1}$ are projectors, and hence, for normalized quantum states $\rho$, $\beta$, $P_U(a|1)=U^{a|1}(\rho)$, $P_V(b|y)=V^{b|y}(\beta)$ satisfy $0 \leq P_U(a|1),P_V(b|y)\leq 1$ . $SWAP$, despite not being positive-semidefinite, is a normalized entanglement witness; hence $0\leq U^{0|0}(\rho)\leq 1$ for all separable states $\rho$. Both $U$ and $V$ thus represent valid deterministic transformations from separable states to classical boxes.

Contracting $U,V,{\cal C}$, we obtain a three-qubit entanglement witness $X$, that we can express as an operator acting on $\C^2\otimes\C^2\otimes\C^2$. Now, consider an optimization over PPT three-qubit states $\rho_{ABC}$, i.e., consider the problem

\begin{align}
&\min\tr(X\rho_{ABC})\nonumber\\
\mbox{such that }&\rho_{ABC},\rho^{T_A}_{ABC},\rho^{T_B}_{ABC},\rho^{T_C}_{ABC}\geq 0, \nonumber\\
&\tr(\rho_{ABC})=1.
\end{align}

This problem can be cast as an SDP; hence we can solve it. The solution is $-0.0721$. This is surprising because neither the CHSH inequality nor the SWAP operator can, by themselves, detect PPT entanglement. Their composition, however, does. So, even if we were not aware of the existence of non-decomposable entanglement witnesses (those which can detect PPT states), we could have derived them from compositional arguments alone.

\section{Conclusion}
\label{conclusion}

We have presented a general method to analyze complex networks, be they classical, quantum or supra-quantum. In essence, our method consists in acting on the many-body system in question with a number of linear transformations---the connectors--- which iteratively coarse-grain the system to one that is small enough to analyze with the existing mathematical tools. While we could relate $m\to 1$ connectors to past literature in Bell nonlocality and entanglement theory, $m\to m'$ connectors seem to be a completely different beast. We showed that connector theory is powerful enough to detect Bell nonlocality (quantum and supraquantum) and entanglement in networks composed of hundreds of sites. Even though we focused on these three areas, we suspect that connector theory will soon find application in other scenarios, for example, to build new dimension witnesses.

Connectors are a natural tool to analyze large, complex many-body systems, and we feel that future research should focus on understanding their mathematical properties. In this regard, our work leaves open important theoretical questions.

One of them is to understand the limitations of the new formalism. Could there be, e.g., entangled tripartite quantum states, undetectable by the composition of a $2\to 1$ connector and a bipartite witness? If not, one wonders how difficult it is in general to find the `right' connectors to detect a particular state or box. The performance of our current numerical methods oscillates between disappointing (it sometimes takes ages to identify the appropriate connectors, even for $m=3$) and excellent ($100$ sites in less than 2 minutes!). Actually, in some scenarios, like \textbf{QUANT-world}, we altogether avoided discussing how to optimize over general $m\to m'$ connectors! 

Another question pertains the practical use in experiments of connector-generated witnesses. Estimating the average value of a witness on a many-body state/box generally requires a number of experiments that scales exponentially with the system size. Is there any way to exploit the tensor network structure of a witness in order to estimate its value with a polynomial number of experiments?

Finally, it is an intriguing idea whether more complicated connectors could be devised by working on a GPT where states are identified with the connectors themselves (\textbf{CONNECTOR-world}), or even with connectors of connectors.

\vspace{10pt}
\noindent\textbf{Acknowledgements}

\noindent We thank S. Pironio and D. Roset for interesting discussions. This work was funded by the Austrian Science fund (FWF) stand-alone project P 30947, the ERC CoG QITBOX, the AXA Chair in Quantum Information Science, the Spanish MINECO (QIBEQI FIS2016-80773-P and Severo Ochoa SEV-2015-0522), Fundaci\'o Cellex, Generalitat de Catalunya (SGR 1381 and CERCA Programme). As part of the \textit{Gravity, Quantum Fields and Information group} at the Albert Einstein Institute, SS acknowledges the support of the provided by the Alexander von Humboldt Foundation and the Federal Ministry for Education and Research through the Sofja Kovalevskaja Award. 

\bibliography{connectBib}

\begin{appendix}
\section{Characterization of the connectors in \textbf{LOC-world}}
\label{appLoc}

Consider a system in \textbf{LOC-world} of the type $(O_1,...,O_m,I_1,...,I_m)$. The state of any such system can be expressed as a convex combination of $\prod_{k=1}^mO_k^{I_k}$ extreme classical boxes. If we order them as $\{P^i(a_1,...,a_m|x_1,...,x_m)\}_i$, then verifying whether a (normalized) box is classical can thus be cast as the LP:

\begin{align}
&\min 0\nonumber\\
\mbox{such that } &P(a_1,...,a_m|x_1,...,x_m)=\nonumber\\
&=\sum_{i}p_iP^i(a_1,...,a_m|x_1,...,x_m),\nonumber\\
&p_i\geq 0, \sum_i p_i=1.
\end{align}
The distribution $\{p_i\}_i$ is sometimes called \emph{local hidden variable model}. For general non-signalling boxes, it does not exist and, when it does, in general it is not unique.

Also due to this finiteness of extreme points, the problem of optimizing over $m\to q$ connectors can be cast as a linear program too. Indeed, let $\{P^i\}_i$ ($\{\tilde{P}^j\}_j$) be the set of extreme points of the output (input) $q$-partite ($m$-partite) \textbf{LOC-world} system, and suppose that, for some fixed tensor $C$, we wished to minimize $W(C)$ over all non-deterministic connectors $W$. The corresponding program would be:

\begin{align}
&\min W(C)\nonumber\\
\mbox{such that } &W(\bar{P}^i)=\sum_j p^i_j\tilde{P}_j,\nonumber\\
&p^i_j\geq 0, \sum_j p^i_j\leq 1,
\label{transLoc}
\end{align}
\noindent where the last condition enforces that the norm of the box does not increase after we apply the connector. To see that any linear functional $W$ satisfying the feasibility conditions transforms local boxes into subnormalized local boxes, note that any initial classical box $P$ admits a decomposition $\sum_i p_iP^i$, with $p_i\geq 0, \sum_ip_i=1$. The result of applying $W$ over such a box is thus the box $\tilde{P}=\sum_{i,j} p_ip^i_j\tilde{P}_j$. Identifying $\mu_j\equiv\sum_ip_ip^i_j$ with our local hidden variable model, we find that $\tilde{P}$ is also Bell-local.

Program (\ref{transLoc}), although correct, can be greatly improved. In the following, we show how to do so by exploiting the lessons learned from the monogamy of non-local correlations. We will do so in three stages. First, we will introduce a convenient notation to deal with no-signalling boxes, that will also be useful to minimize the complexity of LPs like (\ref{transLoc}). A characterization of the dual of the set of non-signalling boxes will follow. Finally, building on the above two results, we present our proposal for linear optmizations over $2\to 1$ connectors.

\subsection{Notation of non-signalling boxes}
\label{notation}
Due to the no-signalling conditions any non-normalized non-signalling box of the form $P(a_1,...,a_k|x_1,...,x_k)$, with $a_i\in\{1,...,d_i\}$, $x_i\in\{1,...,m_i\}$, can be expressed in terms of just $\prod_{i=1}^k(m_i(d_i-1)+1)$ parameters. One way to do so is to adopt what we will call from now on \emph{the abbreviated form} $P(A_1,A_2,...,A_n)$, where $A_i\in \{\emptyset\}\cup\{1,...,d_i-1\}\times\{1,...,m_i\}$. Here $\emptyset$ denotes that the random variable was not measured. That way, e.g., for $a_2\in\{1,...,d_2-1\}$, $P(\emptyset,[a_2,x_2])$ represents the probability that the second party conducted measurement $x_2$ obtained the result $a_2$, i.e., $P(\emptyset,[a_2,x_2])=P_2(a_2|x_2)$.

If we represent the probabilities $P(a_1,...,a_k|x_1,...,x_k)$ ($P(A_1,...,A_k)$) as a vector $\bar{P}$, $\bar{Q}$, there exists a matrix $S$ such that $\bar{P}=S\bar{Q}$.

\subsection{The dual of no-signalling boxes}
\label{NS}

Consider the $k$-partite non-locality scenario $\{(d_i,m_i)\}_{i=1}^k$. In abbreviated form, the corresponding set of (non-normalized) no-signalling distributions is ${\cal B}=\{\bar{q}:S\bar{q}\geq 0\}$, where $\bar{q}\in \R^D$, with $D=\prod_{i=1}^k(m_i(d_i-1)+1)$, and $S$ is the matrix that transforms a box from its abbreviated representation $P(A_1,...,A_k)$ to its standard representation $P(a_1,...,a_k|x_1,...,x_k)$, see Section \ref{notation}. The condition $S\bar{q}\geq 0$ enforces that all the probabilities of the box are non-negative.

The set of positive linear functionals in abbreviated representation is given by the set ${\cal B}'=\{S^T\bar{c}:\bar{c}\geq 0\}$. Indeed, by definition, any $\bar{v}\in L'$ satisfies $\bar{v}\cdot\bar{q}\geq 0$ for all $\bar{q}\in L$, and so the dual set of ${\cal B}$ contains ${\cal B}'$. It rests to show that any vector outside ${\cal B}'$ cannot belong to the dual of ${\cal B}$. First note that, for any $\bar{q}\not\in {\cal B}$, there exists $\bar{v}\in {\cal B}'$ such that $\bar{v}\cdot\bar{q}< 0$ (take, e.g., $\bar{v}=S^T\bar{c}$, with $c_j=\Theta(-(S^T\bar{q})_j)$). Now, let $\bar{w}\not\in {\cal B}'$. By the Separation theorem there exists $\bar{q}$ such that $\bar{v}\cdot\bar{q}\geq 0$ for all $\bar{v}\in {\cal B}'$, and $\bar{w}\cdot\bar{q}<0$. The first condition implies that $\bar{q}\in {\cal B}$, and so the second condition implies that $\bar{w}$ is not in the dual of ${\cal B}$.

With the formulation above, it is clear that linear optimizations over the set of positive functionals of no-signalling boxes can be carried out via linear programming \cite{lp}.

\subsection{Faster codes for optimization over $2\to 1$ connectors}
\label{appLocTrick}

First, we will define a (non-normalized) local box in a non-standard way.

\begin{defin}
The probabilities $\{P(a,b|x,y):x=1,...,n_A;y=1,...,n_B;a=1,...,d_A;b=1,...,d_B$ define a local box iff there exist $\{P(a,b_1,...,b_{n_B}|x)\}$ such that

\begin{align}
&P(a,b_y|x,y)=\sum_{b_z:z\not=y} P(a,b_1,...,b_{n_B}|x),\nonumber\\
&\sum_a P(a,b_1,...,b_{n_B}|x)=P(b_1,...,b_{n_B}),\nonumber\\
&P(a,b_1,...,b_{n_B}|x)\geq 0.
\label{extended}
\end{align}
\end{defin}

That this definition implies bipartite locality can be seen by noting that the variables $b_1,...,b_{n_B}$ play the role of local hidden variables in the decomposition above. Conversely, let $P(a,b|x,y)=\sum_\lambda P(\lambda)P_A(a|x,\lambda)P_B(b|y,\lambda)$. Then one can verify that $P(a,b_1,...,b_{n_B}|x)\equiv \sum_{\lambda}P(\lambda)P_A(a|x,\lambda)P_B(b_1|1,\lambda)...P_B(b_{n_B}|n_B,\lambda)$ satisfies the conditions in (\ref{extended}). From now on, we will refer to the object $P(a,b_1,...,b_{n_B}|x)$ as an \emph{extended box}, and represent it by the vector of probabilities $\bar{P}$.

Note that we can regard an extended box as a no-signalling box where all the parties but the first have just one input. Therefore, we can represent extended boxes in abbreviated form, as a vector of probabilities $\bar{Q}=P(A,B_1,...,B_{n_B})$, with $A\in \{\emptyset\}\cup\{1,...,d_A-1\}\times\{1,...,n_A\}$, $B_i\in \{\emptyset\}\cup\{1,...,d_B-1\}$. Let $S$ be the matrix that effects the transformation $S\bar{Q}=\bar{P}$. In abbreviated form, the set of non-normalized local boxes is thus described by $L\equiv\{\bar{q}:S\bar{q}\geq 0\}$. As proven in Section \ref{NS}, the dual of this set, i.e., the set of vectors $\bar{v}$ such that $\bar{v}\cdot \bar{q}\geq 0$ for all $\bar{q}\in L$, corresponds to the set of vectors $L'\equiv \{S^T\bar{c}:\bar{c}\geq 0\}$.

Any positive linear functional over local boxes must remain positive if we embed it into the space of extended boxes. This provides us with a computationally efficient characterization of the set of positive Bell functionals.

\begin{prop}
$\{U(A,B)\}$ is a positive functional over the set of local boxes iff there exists a vector $\bar{c}\geq 0$ such that

\begin{align}
&S^T\bar{c}(A,B_1,...,B_{n_B})=0,\mbox{ if }\exists i,j: i\not=j, B_i,B_j\not=\emptyset\nonumber\\
&S^T\bar{c}(A,B_1,...,B_{n_B})=U(A,B),\mbox{ if } B_y=B,B_z=\emptyset, \forall z\not=y.
\end{align}

\end{prop}

With this formulation, one can carry out linear optimizations over the set of positive functionals of local boxes via linear programming \cite{lp}. The computational cost will be bearable provided that $n_B$ is not very large. $n_A$ can take high values, though.

\section{$m\to 1$ connectors for quantum boxes}
\label{appQuant}
The set of quantum boxes can be formulated as the set of all boxes of the form 

\be
P(a_1,...,a_k|x_1,...,x_k)=\tr(\rho E^1_{a_1,x_1}\otimes...\otimes E^k_{a_k,x_k}),
\label{Qbox}
\ee
\noindent where $\rho$ is a positive semidefinite matrix with $\tr(\rho)\leq 1$ and $\{E_k^{a,x}\}$ satisfy 

\begin{align}
&(E_j^{a,x})^\dagger=(E_j^{a,x})^2=E_j^{a,x},\nonumber\\
&\sum_{a}E_j^{a,x}=\id_j.
\label{relations}
\end{align}

In order to derive positive functionals for quantum boxes, we will rely on non-commutative polynomial optimization theory \cite{npa, npa2, npo}. Let $X_0,X_1,...,X_n$ be a number of Hermitian operators acting on the same Hilbert space, with $X_0=\id$ and let $\rho$ be a normalized quantum state. The $s^{th}$-order moment matrix $\Gamma$ of this system is the matrix whose rows are columns are labeled by words of the alphabet $\{0,...,n\}$ of length $s$ or smaller, and whose entries are given by

\be
\Gamma_{\bar{i},\bar{j}}=\tr(\rho X_{\bar{i}}^\dagger X_{\bar{j}}),
\ee
\noindent where, for any word $\bar{i}$ of length $t$, $X_{\bar{i}}\equiv X_{i_1}...X_{i_t}$. It can be verified that all moment matrices are positive semidefinite \cite{npo}.

Now, consider the moment matrices defined by the quantum state $\rho$ and operators $\{\id_{1,...,j-1}\otimes E^{a,x}_j\otimes \id_{j+1,...,k}\}$ generating our quantum box (\ref{Qbox}). It is immediate to see that, for moment matrices $\Gamma$ of high enough order, there exist matrices $F_{A_1,...,A_k}$ such that $P(A_1,...A_k)=\tr(\Gamma F_{A_1,...,A_k})$. It is also easy to see that the moment matrices of unnormalized quantum boxes are subject to non-trivial linear constraints \cite{npa,npa2}. That is, there exists a set of matrices $\{G_j\}_j$ such that any moment matrix $\Gamma$ can be expressed as $\Gamma=\sum_j c_j G_j$, for some choice of coefficients $\{c_j\}_j$.

From the above, it follows that a sufficient condition for $w(A_1,...,A_k)$ to be a positive linear functional over the set of quantum boxes is that there exists a matrix $Z\geq 0$ such that 

\be
\tr\left(\sum_{A_1,...,A_k}w(A_1,...,A_k)F_{A_1,...,A_k}G_j\right)=\tr(ZG_j),
\ee
\noindent for all $j$.

Indeed, note that 

\begin{align}
&\sum_{A_1,...,A_k}w(A_1,...,A_k)P(A_1,...,A_k)=\nonumber\\
&\tr\left(\sum_{A_1,...,A_k}w(A_1,...,A_k)F_{A_1,...,A_k}\Gamma\right)=\nonumber\\
&\tr\left(Z\Gamma\right)\geq 0.
\label{Qcriterion}
\end{align}
Here the second equality stems from the fact that $\Gamma=\sum_j c_j G_j$. The inequality holds because both $\Gamma$ and $Z$ are positive semidefinite matrices.

\section{$m\to m'$ connectors for \textbf{SEP-world}}
\label{appSep}
First we will explain how to optimize over $m\to 1$ connectors which transform multipartite separable states with Hilbert space dimension $d_1\times d_2\times...d_m$ into a quantum state in dimension $d_{m+1}$. We will use the Choi-Jamiolkowski notation to represent connectors, i.e., each connector $\Omega$ will be identified with an $m+1$-partite operator $W_{1,...,m+1}$ such that $\Omega(\rho_{1,..,m})=\tr_{1,...,m}\{W_{1,...,m+1}(\rho_{1,...,m}\otimes \id_{m+1})\}$.

Call $\H_{sym}^{k,d}$ the symmetric subspace of $k$ identical particles of dimension $d$. Then $S^k$ is the set of $m+1$-partite states $\sigma$ such that there exists a $km+1$-partite state $\beta_{1,...,1,2,...,2,...,m,....,m,m+1}\in B(\H_{sym}^{k,d_1})\otimes...B(\H_{sym}^{k,d_m})\otimes B(\C^{d_{m+1}})$, satisfying

\begin{enumerate}
\item
$\beta\geq 0$.
\item
$\beta^{T_A}\geq 0$ has a Positive Partial Transpose (PPT) \cite{peres} for all bipartitions $A$ of the systems $\{1,...,1,2,...,2,m,...,m,m+1\}$.
\item
$\tr_{\Lambda}(\beta)=\sigma$, where $\Lambda$ denotes any set of indices with $k-1$ $1$'s, $2$'s,...,$m$'s.
\item
$\beta\Pi^k_{sym}=\beta$.
\end{enumerate}
Here $\Pi^{k}_{sym}$ denotes the tensor product of the symmetric projectors on the spaces $(\C^{d_j})^{\otimes k}$, for $j=1,...,m$, times the identity on $\C^{d_{m+1}}$.

Intuitively, $\beta$ represents the PPT state of an ensemble of $m$ groups of $k$ identical Bosons, plus a third particle labeled $m+1$. Any $\beta\in B(\H_{sym}^{k,d_1})\otimes...B(\H_{sym}^{k,d_m})\otimes B(\C^{d_{m+1}})$ satisfying the properties above is called a Bose-symmetric PPT $k$-extension of $\sigma$.

\noindent As proven in \cite{DPS3}, any separable state admits a Bose-symmetric PPT $k$-extension for all $k$. Indeed, let $\sigma=\sum_{i} p_i\bigotimes_{j=1}^{m+1}\proj{u^j_i}$. Then it can be verified that the state $\beta=\sum_{i} p_i\bigotimes_{j=1}^{m}\proj{u^j_i}^{\otimes k}\otimes \proj{u^{m+1}_i}$ satisfies the above constraints. Most importantly, the limiting set $\lim_{k\to\infty}S^k$ is the set of fully separable states \cite{DPS3}.

As explained in the main text, rather than over general entanglement witnesses, we will conduct optimizations over a subset thereof. More precisely, we will consider a set ${\cal W}_m^k$ of multipartite operators $W_{1,...,m+1}$ such that $\tr\{W\sigma\}\geq 0$ for all states $\sigma\in S^k$. 

This set is composed by operators $W_{1,...,m+1}$ such that 

\be
\Pi^{k}_{sym}(W_{1,...,m+1}\otimes \id_{d_1}^{\otimes k}\otimes ...\otimes \id_{d_m}^{\otimes m})\Pi^{k}_{sym}=\sum_{A}V_A^{T_A},
\label{condiMAP}
\ee
\noindent with the sum on the left running over all bipartitions $A$ of the $km+1$ parties and $V_A\geq 0$, for all partitions $A$. 

We will next prove that any such operator satisfies $\tr(W_{1,...,m+1}\sigma)\geq 0$ for all states $\sigma$ admitting a Bose-symmetric PPT $k$-extension on systems $1,2,...,m$. this implies, in particular, that $W$ is an entanglement witness.

Let then $\sigma$ admit a Bose-symmetric PPT $k$-extension $\beta_{1,...,1,2,...,2,...,m,...,m,m+1}$. Then we have that

\begin{align}
&\tr(W\sigma)=\tr\{(W\otimes \id_{d_1}^{\otimes k}\otimes ...\otimes \id_{d_m}^{\otimes m})\beta\}=\nonumber\\
&\tr\{\Pi^{k}_{sym}(W\otimes \id_{d_1}^{\otimes k}\otimes ...\otimes \id_{d_m}^{\otimes m})\Pi^{k}_{sym}\beta\}=\nonumber\\
&\tr\{\sum_AV^{T_A}_A\beta\}=\sum_A\tr\{V_A\beta^{T_A}\}\geq 0.
\end{align}
\noindent Here the first equality follows from the fact that $\beta$ is an extension; the second, from it living in the symmetric subspace; and the third, from eq. (\ref{condiMAP}). The last inequality follows from the fact that $\beta$ is PPT and that $V_A\geq 0$ for all bipartitions $A$.

It hence follows that any map $\Omega$ satisfying the SDP conditions:

\begin{enumerate}
\item
$W_\Omega\in {\cal W}^k_m$,
\item
$\id_{1,...,m}-\tr_{m+1}(W_\Omega) \in {\cal W}^k_{m-1}$,
\end{enumerate}
\noindent is a $m\to 1$ connector in \textbf{SEP-world}. Clearly, linear optimizations over this set can be cast as an SDP.

The DPS hierarchy also provides us with tools to define SDP ans\"atze of $m\to m'$ connectors. In \cite{NOP}, it is shown that, for any state $\rho_{1,...,m'}$ admitting a Bose-symmetric extension (note that the PPT condition is not necessary), the state $\tilde{\Omega}(\rho)\equiv\bigotimes_{j=1}^{m'-1}\Omega_j^{d_j,k}(\rho)$ is separable. Here $\Omega^{d,k}$ denotes the partially depolarizing channel 

\be
\Omega^{d,k}(\sigma)=\frac{k}{k+d}\sigma+\frac{d}{k+d}\frac{\id}{d}.
\ee

It follows that any $m\to 1$ connector with output in $B(\H_{sym}^{k,d'_1}\otimes...\otimes\H_{sym}^{k,d'_{m'+1}}\otimes\C^{d_{m'}})$ can be transformed into a $m\to m'$ connector just by tracing out the extra systems and applying $\tilde{\Omega}$ at the output.

Let us finish with a trick to optimize over $m\to 2$ connectors when the output is a $\C^2\otimes\C^2$ or a $\C^2\otimes\C^3$ system. We again start from an $m\to 1$ connector $W$, with output spaces $A_{m+1}$, $B_{m+1}$. The key idea is to enforce that both $W_\Omega$ and $W_\Omega^{T_{A_{m+1}}}$ are entanglement witnesses. If that is the case, then the output of the map will be a PPT state, and so, by \cite{horo}, a separable state. Imposing that $W_\Omega, W_\Omega^{T_{A_{m+1}}}\in{\cal W}^k_m$ is again an SDP.

\section{Non-signalling boxes admitting an MPS decomposition}
\label{appNSBoxes}

The tensors $\Lambda^{[k]}_{a_k,x_k}$ of the general Svetlichny box are given by:

\begin{align}
&\Lambda^{[1]}_{a,x}=\frac{1}{2}\bra{x}\bra{x}\bra{0}\bra{a},\nonumber\\
&\Lambda^{[k]}_{a,x}=\frac{1}{2}\id_2\otimes\sum_{y=0,1}\ket{y}\bra{x}\otimes X^{yx}\otimes X^{a},\mbox{ for }1<k<m,\nonumber\\
&\Lambda^{[m]}_{a,x}=\sum_{y,z,s=0,1}\ket{z}\ket{y}\ket{s}(X^{x(y+z)+a}\ket{s}).
\end{align}

\noindent Here $X$ denotes the Pauli matrix $\left(\begin{array}{cc}0&1\\1&0\end{array}\right)$.

The result can be proved by induction, but, to get an intuition on the construction, consider the vector $\bra{\psi_k}\equiv \Lambda^{[1]}_{a_1,x_1}\Lambda^{[2]}_{a_2,x_2}...\Lambda^{[k]}_{a_{k},x_{k}}$. It can be verified that $\bra{\psi_k}=\frac{1}{2^k}\bra{x_1}\bra{x_{k}}\bra{x_1x_2\oplus...\oplus x_{k-1}x_{k}}\bra{a_1\oplus...\oplus a_k}$. That is: the first qubit register contains a copy of the value of $x_1$; the second, the value of $x_{k}$; the third, the part of $f(x_1,...,x_m)$ computed so far; and the last one, the part of $a_1\oplus...\oplus a_m$ computed so far.

The tensors defining box (\ref{consecBox}) are given by:

\begin{align}
&\Lambda^{[1]}_{a,x}=\frac{1}{2}\bra{x}\bra{a},\nonumber\\
&\Lambda^{[k]}_{a,x}=\frac{1}{2}M_{x,r}\otimes X^a,\mbox{ for }1<k<m,\nonumber\\
&\Lambda^{[m]}_{a,x}=(M_{x,r}\otimes X^a)\left(\ket{r}\ket{1}+\sum_{j=0}^{r-1}\ket{j}\ket{0}\right),
\end{align}
\noindent where $M_{x,r}=\proj{r}+\sum_{j=0}^{r-1}\ket{j}\bra{x(j+1)}$. In this case, the first register has $r+1$ levels ($\ket{0},...,\ket{r}$), and it represents a counter. The second register is a qubit carrying the sum modulo $2$ of the outputs.

Finally, it can be verified that an MPS representation for box (\ref{majBox}) is given by the matrices:

\begin{align}
&\Lambda^{[1]}_{a,x}=\frac{1}{2}\bra{x}\bra{a},\nonumber\\
&\Lambda^{[k]}_{a,x}=\frac{1}{2}\tilde{M}_{x,m}\otimes X^a,\mbox{ for }1<k<m,\nonumber\\
&\Lambda^{[m]}_{a,x}=(\tilde{M}_{x,m}\otimes X^a)\left(\ket{\left\lceil \frac{m}{2}\right\rceil}\ket{1}+\sum_{j=0}^{\lceil \frac{m}{2}\rceil}\ket{j}\ket{0}\right),
\end{align}
\noindent where $\tilde{M}_{x,m}=\ket{\lceil \frac{m}{2}\rceil}\bra{\lceil \frac{m}{2}\rceil}+\sum_{j=0}^{\lceil \frac{m}{2}\rceil-1}\ket{j}\bra{j+x}$. This time, the first register has $\lceil \frac{m}{2}\rceil+1$ levels ($\ket{0},...,\ket{\lceil \frac{m}{2}\rceil}$).

\section{Non-trivial $2\to2$ connector}
\label{appDisEnt}

Consider the $[2,2]\to [2,2]$ connector $W$ given by the matrix:

\be
W=\left(\begin{array}{ccccccccc}
1 & 0  & 0 & 0 & 0 & 0 & 0 & 0 & 0\\
\frac{1}{4} & 0 & 0 & \frac{1}{4} & \frac{1}{2} &   -\frac{1}{4} &   \frac{1}{2} &  -\frac{1}{2} & \frac{1}{4}\\
\frac{1}{2} &  -\frac{1}{4} & \frac{1}{2} & \frac{1}{2} &  -\frac{1}{4} &  -\frac{1}{2} & 0 & \frac{1}{4} & -\frac{3}{4}\\
\frac{3}{4} &  -\frac{1}{2} &  0 & \frac{1}{4} & \frac{1}{2} &  -\frac{1}{4} &  -\frac{1}{2} &  \frac{1}{4} & \frac{1}{2}\\
\frac{1}{4} & 0 &  -\frac{1}{4} & \frac{1}{4} &  \frac{1}{2} &  -\frac{1}{4} & 0  &  -\frac{1}{4} & \frac{3}{4}\\
\frac{1}{2} & -\frac{1}{2} &  \frac{1}{4} &  \frac{1}{2} & 0 &  -\frac{1}{2} &  -\frac{1}{2} &  \frac{1}{2} &  -\frac{1}{4}\\
\frac{3}{4} & 0 & 0 &  -\frac{1}{4} & -\frac{1}{2} & \frac{1}{2} &  -\frac{1}{4} &  \frac{1}{4} & \frac{1}{4}\\
\frac{1}{4} & 0 &  -\frac{1}{4} & -\frac{1}{4} & 0& \frac{1}{2} & \frac{1}{4} & -\frac{1}{4} & \frac{1}{2}\\
\frac{1}{4} & 0 & \frac{1}{2} & \frac{1}{4} & -\frac{1}{2} & 0 & -\frac{1}{4} & \frac{1}{4} & -\frac{1}{2}\end{array}\right)
\ee

We are using the abreviated notation, i.e., both the input and output distributions $P(A,B), Q(C,D)$ are represented as $9$-entry vectors, with $Q=W\dot P$. It can be verified that $W$, acting over any deterministic point, generates a local bipartite box. However, $W\cdot\bar{P}$ generates a ``box'' with negative probabilities, when $\bar{P}$ corresponds to the one of the variants of the PR-box, $P(a,b|x,y)=\frac{1}{2}\delta(\bar{a}\oplus\bar{b},\bar{x}\bar{y})$ \cite{pr}, where $\delta(s,t)$ denotes the Kronecker delta. $W$ is thus not a wiring. Furthermore, it can be verified, using semidefinite programming, that

\be
\min_{W'}\|W-W'\|_\infty\approx 0.078,
\ee
\noindent where the minimum is taken over all non-deterministic transformations $W'$ (in matrix form) which can be factored out as shown in \fref{disentPic}.

\end{appendix}

\end{document}